\documentclass{imammb}

\jno{dqnxxx}

\usepackage{natbib}

\usepackage{graphicx}
\usepackage{amsmath,amsthm,bm,mathrsfs}
\usepackage{amssymb}
\usepackage{subcaption, floatrow}
\usepackage[backref=true,pagebackref=true,hyperindex=true,breaklinks=true,colorlinks=true,urlcolor=green,bookmarks=true,bookmarksopen=false,citecolor=blue,linkcolor=blue]{hyperref}
\usepackage{doi}
\usepackage[percent]{overpic}
\usepackage{multirow}
\usepackage{xcolor}
\usepackage{cancel}
\usepackage{mathtools}
\usepackage{easy-todo}

\newcommand{\hide}[1]{}

\newcommand{\red}[1]{#1}

\newcommand{\rev}[1]{#1}
\newcommand{\revm}[1]{#1}

\usepackage[T1]{fontenc}
\usepackage{accents}
\DeclareMathAlphabet\mathbfcal{OMS}{cmsy}{b}{n}
\renewcommand{\@}{\partial}
\renewcommand{\vec}[1]{\mathbf{#1}}

\renewcommand{\d}{\mathrm{d}}

\newcommand{\df}[2]{\frac{\d #1}{\d #2}}

\newcommand{\eq}[1]{\eqref{#1}}

\newcommand{\notcolor}{black}
\newcommand{\+}[3]{\def#1{{\color{\notcolor}#2}}}
\+{\t}{t}{original (slow) time}
\+{\T}{\tau}{fast time, $\t/(\epsone\epstwo)$}
\+{\BCL}{\beta}{the temporal period}
\+{\B}{\beta}{the temporal period}
\+{\A}{\alpha}{APD - asymptotic expression}
\+{\D}{\delta}{DI  - asymptotic expression}
\+{\u}{v}{excitation variable in Tikhonov system}
\+{\v}{w}{recovery variable in Tikhonov system}
\+{\w}{w}{recovery variable in Tikhonov system}
\+{\a}{a}{parameter a}
\+{\c}{b}{parameter c}
\+{\fu}{f}{rate of change of $\u$ variable(s) without $\eps$}
\+{\fv}{g}{rate of change of $\v$ variable(s)}
\+{\fy}{g}{rate of change of $\u$ variable(s) without $\eps$}
\+{\fw}{g}{rate of change of $\u$ variable(s) without $\eps$}
\+{\f}{f}{rate of change of $\u$ variable(s) without $\eps$}
\+{\g}{g}{rate of change of $\v$ variable(s)}
\+{\ath}{a}{threshold parameter in McKean model}
\+{\eps}{\varepsilon}{the small parameter of the main embedding}
\+{\uroot}{\mathcal{V}}{The three roots of $\fu=0$}
\+{\umins}{\u_\text{dia}}{left piece of $\u$-nulcline}
\+{\uplus}{\u_\text{sys}}{right piece of $\u$-nulcline}
\+{\uth}{\u_\text{thr}}{thershold piece of $\u$-nulcline}
\+{\vnot}{\v_\B}{initial condition for $\v$}
\+{\Mmin}{M_\text{min}}{minimum fold point}
\+{\Mmax}{M_\text{max}}{maximum fold point}
\+{\wB}{\w_\B}{initial condition}
\+{\wB}{\w_\B}{initial condition}
\+{\anew}{\tilde{\a}}{1-a}
\+{\cnew}{\tilde{\c}}{1+c}
\+{\Aexp}{A}{APD before drug}
\+{\dAexp}{\Delta A}{DAPD after drug}
\+{\Bexp}{B}{BCL in experiment 2Hz}
\+{\Dexp}{D}{DI before drug}
\+{\be}{\gamma}{drug action scaling factor}
\+{\beexp}{\Gamma}{drug concentration in experiment}
\+{\De}{\mathcal{D}}{experimental dataset}
\+{\Pe}{\mathcal{P}}{dataset of model parameters}
\newcommand{\APDn}{APD$_{90}$}
\newcommand{\dAPDn}{$\Delta$APD$_{90}$}
\newcommand{\APD}{APD}
\newcommand{\dAPD}{$\Delta$APD}
\newcommand{\myrightleftarrows}[1]{\mathrel{\substack{\xrightarrow{#1} \\[-.9ex] \xleftarrow{#1}}}}

\numberwithin{equation}{section}

\allowdisplaybreaks
\begin{document}
\title{Phenomenological analysis of \rev{simple} ion channel block in large populations of uncoupled cardiomyocytes}

\renewcommand*{\thefootnote}{\fnsymbol{footnote}}
\author{
{\sc Radostin D.~Simitev\footnote{Corresponding author. Email: \href{mailto:Radostin.Simitev@glasgow.ac.uk}{Radostin.Simitev@glasgow.ac.uk}. OrcID: \href{https://orcid.org/0000-0002-2207-5789}{orcid.org/0000-0002-2207-5789}.}, Antesar Al Dawoud}\\[2pt]
{\small School of Mathematics \& Statistics, University of Glasgow,
Glasgow G12 8QQ, UK}\\[6pt]
{\sc Muhamad H.N. Aziz}\\[2pt]
{\small Institute of Mathematical Sciences, Universiti Malaya, 50603
K. Lumpur, Malaysia }\\[6pt]
{\sc Rachel Myles and Godfrey L.~Smith}\\[2pt]
{\small Institute of Cardiovascular \& Medical Sciences, University of
Glasgow, Glasgow G12 8TA UK}\\[6pt]
}
\date{\today}

\pagestyle{headings}
\maketitle

\markboth{\rm R.~SIMITEV \textit{ET AL.}}{\rm MODEL OF ION CHANNEL
BLOCK IN LARGE CARDIOMYOCYTE POPULATIONS}


\begin{abstract}
{
\looseness=-1
Current understanding of arrhythmia mechanisms and design of
anti-arrhythmic drug therapies hinges on the assumption that myocytes
from the same region of a single heart have  similar, if not  
identical, action potential waveforms and drug responses. On the contrary, recent
experiments reveal significant heterogeneity
in uncoupled healthy myocytes both from different hearts as well as
from identical regions within a single heart.
In this work, a methodology is developed for quantifying the
individual electrophysiological properties of large numbers of uncoupled
cardiomyocytes under ion channel block in terms of the parameters values of a conceptual fast-slow
model of electrical excitability. The approach is applied
to a population of nearly 500 rabbit ventricular myocytes for which
action potential duration (APD) before and after the application of
the drug nifedipine was experimentally measured (Lachaud et al., 2022,
Cardiovasc. Res.). 
To this end, drug action is represented by a multiplicative factor to an
effective ion conductance, a closed form
asymptotic expression for \APD{} is derived and inverted to
determine model parameters as functions of \APD{} 
and \dAPD{} (drug-induced change in \APD{}) for each myocyte.
Two free protocol-related quantities are calibrated to experiment
using an adaptive-domain procedure based on an original assumption of
optimal excitability. 
The explicit \APD{} expression and the resulting set of model
parameter  values
allow (a) direct evaluation of conditions necessary to maintain
fixed \APD{} or \dAPD{}, (b) predictions of
the proportion of cells remaining excitable after drug application, (c)
predictions of stimulus period  dependency and (d) predictions of 
dose-response curves, the latter being in agreement with additional
experimental data.
}{asymptotics of cellular electrical excitability, action potential duration, drug response
}
\end{abstract}

\section{Introduction}
\looseness=-1
The heart pumps blood due to coordinated contraction of approximately
50 million of
individual cardiac cells. Contraction of each
myocyte is triggered by the excitation of electrical 
impulses known as transmembrane action potentials (AP),
e.g.~\citep{Bers2001}. When disease, 
inherited disorders or environmental factors prolong or shorten the
duration of the action potential 
the heart becomes vulnerable to arrythmias,
electrical instabilities, that may rapidly deteriorate to cause fatal
deficiency in cardiac output \citep{Anumonwo2015,Tse2016}. Thus, there is  strong impetus
to develop anti-arrythmic drugs that can control action potential
duration and restore it to norm  \citep{Darbar2018}.
The action
potential duration (\APD) and its change under drug 
action (\dAPD) are, therefore, primary biomarkers used to guide
the design of anti-arrythmic drugs and quantify their
pharmacodynamics \citep{Corrias2010}.

Novel optics-based techniques in cardiac electrophysiology
have now made it possible to design high-throughput
systems capable of measuring \APD{} and other secondary AP
waveform biomarkers at rates of up to 200 cells/hr \citep{Warren2010,Lachaud2018,Mllenbroich2021}. 
In our recent study \citep{Lachaud2022}, AP characteristics of nearly
500 uncoupled cardiomyocytes were measured using voltage-sensitive
fluorescent dyes. The cells were isolated from well-defined 
regions of the left ventricular wall of 12 rabbit hearts
and \APD{} values were taken from the same cells before and after
specific ion channel inhibition with  
two different drugs.
An unexpectedly large variability in the values of the action potential
duration at 90\% repolarization (\APDn) of the uncoupled 
healthy cardiomyocytes was measured before the addition of the drugs. Specifically, at stimulation rate of 2 Hz the
inter-quartile range of \APDn{} was 40 to 50 ms with median 
value of 250 ms. This variation was not due to cell dissociation
damage and it is considerably larger than 
regional endo-epicardial and apical-basal differences in median \APDn{}
from single hearts, as well as larger than differences  in
median \APDn{} measured between the individual hearts.
Measurements after inhibition of the IK(r) ionic
current by 30 nM of dofetilide  and after inhibition of the ICa(L)
current by 1 $\mu$M of nifedipine both showed that individual cells with near
identical baseline values of \APDn{} produce a wide range of
different \dAPDn{} values. The latter result demonstrates starkly
that \APD{} alone does not characterise or 
uniquely determine the electrophysiological response of myocytes to
drugs, as often assumed.
\rev{Measurements of additional, mutually-independent biomarkers are
required increase the accuracy of \dAPDn{} estimation irrespective of
whether a data-driven or physics/physiology-based approach is employed.}
It is the goal of our work to interpret these findings in the light
of a conceptual mathematical model as discussed next.

To understand this significant intrinsic variability, action potential waveforms, which are of ionic
origin \citep{Pandit2018}, must be related to their underlying
electrophysiological characteristics, including  
the conductances and kinetic parameters of the ion channels, exchangers
and pumps of each individual myocyte.
However, high-throughput patch-clamping of ionic currents in large
numbers of cells (more than 100) are 
currently not feasible, and clamping myocytes already used with
voltage-fluorescent dyes is an even bigger technical
challenge. Mathematical ionic-current models of the 
AP waveform provide a valuable alternative \citep{Clayton2011,Davies2016,Clayton2020}. 
Following \citep{Gemmell2016,Muszkiewicz2016}, a rejection sampling procedure
was used in \citep{Lachaud2022} for this purpose. The modelling procedure consisted of
(a) selecting the detailed ionic current model of \cite{Shannon2004}
as a mathematical representation of the rabbit myocytes,
(b) a parameter sensitivity analysis to determine the ionic
conductances in the model that most strongly affecting the AP waveform
followed by (c) their random variation to generate a model 50,000 
variants, and finally (d) a calibration of the ensemble
by rejecting model variants that fell outside the ranges and the
histogram distribution of the experimentally measured \APDn{}
values. However, the \cite{Shannon2004} model variants calibrated in
this way were not cell-specific, the population was not unique and 
less sensitive parameters remained at baseline values, for reasons
outlined below.

While offering valuable insight and being routinely employed to
interpret experimental findings, extrapolate animal data
to human system context and test novel hypotheses, detailed ionic
models such as that of \cite{Shannon2004} are prohibitively complicated \citep{Sigg2010}.
For example, the latter model consists of 45 ordinary
differential equations and includes 177 model parameters. 
Many of these parameters and equations are poorly constrained, some even
redundant, because such models are typically developed by extending and re-using components
from earlier models, as advocated by large international initiatives
like the Physiome Project \citep{Bassingthwaighte2000} and the CellML Project \citep{Miller2010}.
For instance, the modern human ventricular models
of \cite{tenTusscher2004} and \cite{Iyer2004} include parameters inherited from
studies in at least 9 different species over a range of 6 different
temperatures \citep{Niederer2009}, and this is likely true for the
model of \cite{Shannon2004}, as well.
Despite intensive research effort expended to estimate their parameter
uncertainty \citep{Clayton2020}, calibrate models to identifiable and
reliable experimental protocols \citep{Whittaker2020,Clerx2019} and increase
their reproducibility \citep{Cooper2016,Johnstone2016}, detailed
cardiac cell models remain difficult to adapt to situations to which
they have not been fitted \citep{Wilhelms-2013}. Most 
importantly, detailed cardiac models are becoming increasingly
difficult for causal inference \citep{Biktashev-2008}, as also evidenced
by \rev{the need to resort to} rejection sampling procedures such as described above
in relation to the analysis of the data of \cite{Lachaud2022}.

With this motivation, the aim of our study is to employ a simple
phenomenological model of the transmembrane action potential in order to
formulate a mathematical description of the experimental procedure
of \cite{Lachaud2022}.
\rev{
A variety of simplified phenomenological models of the cardiac action
potential exist \citep{FitzHugh1961,Nagumo1962,Aliev1996,Mitchell2003},
to list a few.}
The model of \cite{McKean1970},
featuring piece-wise linear kinetics and only three intrinsic
parameters will be used here, being arguably the simplest such model.
The \cite{McKean1970} equations allow exact solution for the
action potential waveform in closed form. Here, a phase-space
analysis will be used to derive an even simpler invertible asymptotic
relationship between \APD{} and the parameters of the model
with the goal of illustrating explicitly the geometric behaviour of
these quantities. A natural calibration of the model 
to the experimental data of \citep{Lachaud2022} will be proposed that
will allow to determine uniquely the individualised values of
the \cite{McKean1970} model parameters corresponding to each rabbit
myocyte used in experiments. Admittedly,  the \cite{McKean1970} model
and its parameters have no direct correspondence to myocyte
electrophysiological structures and processes such as ion channel
conductances and kinetic parameters of transmembrane currents. However, the
analysis is valuable as it provides an mathematically well-defined
test case that can be used to conceptualise experiments and validate
\rev{other} parameter inference and \rev{data-driven} approaches such as rejection
sampling and calibration \citep{Muszkiewicz2016,Gemmell2016,Whittaker2020}, inverse 
regression \citep{Sobie2009,Sarkar2010}, machine learning \citep{Feeny2020,Trayanova2021},
Gaussian and Bayesian emulation \citep{Coveney2020,Coveney2021} and 
multi-objective optimisation \citep{Pouranbarani2019} of detailed
ionic AP models.

\rev{For completeness, we mention data-driven models as an alternative
strategy for studying the intrinsic variability of action potential
waveforms. A data-driven model is a statistical model for prediction
of a target quantity as a function of observed features without
recourse to an intermediate ``first-principles'' model. 
Examples include regression and neural-network models; \cite{Hastie2009} provides an
extensive general overview while specific applications of data-driven approaches
to cardiac cellular excitability were already cited above.
Typically, data-driven models are relatively accurate but are less
interpretable, computationally efficient and generalizable than
physics/physiology-based models. The development of purely data-driven
approaches is at present limited by experimental factors. For
instance, measurements of Trise, APD$_{50}$, APD$_{30}$, and
triangulation index are also reported in \citep{Lachaud2022}, but the 
uncertainty in measuring TRise is significant, while biomarkers
APD$_{XX}$ are strongly correlated with each other. 
} 

\section{Asymptotic approximation of \APD{} in the McKean model}

\subsection{The McKean equations}
\looseness=-1
To interpret the results of \cite{Lachaud2022}, we consider,
as a phenomenological model of the action potential
of uncoupled rabbit ventricular myocytes,
the following planar system of first-order ordinary differential equations
\begin{subequations}
\label{McKean}
\begin{align}
\label{McKean-Eq1}
& \df{\u}{\t} = \eps^{-1} \f(\u,\w;\a,\c), &\f(\u,\w; \a,\c)&:=-\Big(\u-H(\u-\a)+\w\Big),\\
\label{McKean-Eq2}
& \df{\w}{\t} = \g( \u,\w; \a,\c), & \g(\u,\w; \a,\c)&:= \u-\c\w.
\end{align}
\end{subequations}
Here $\u$ and $\w$ are dynamical variables of time $\t$, interpreted as
the myocyte trans-sarcolemmal voltage potential and an effective gating
variable, respectively, $H(x)$ denotes the Heaviside step function,
and $\a$, $\c$ and $\eps$ are model parameters with $0<\eps\ll1$.
In the experiments of \cite{Lachaud2022}, myocytes were stimulated at 2 Hz using a 2 ms
duration voltage pulse at 1.5 threshold via carbon electrodes. To
mimic this stimulation protocol,
we complement
equations \eqref{McKean} by the initial conditions  
\begin{subequations}
\begin{gather}
\label{stimvolt1}
\u(0)=\u_\text{stim}>\a,~~~~\w(0)=\w_0,
\end{gather}
and then advance via a sequence of initial value problems on time
intervals $\t\in\big(k\B,(k+1)\B\big]$, $k\in\mathbb{N}$ 
with duration $\B$ (basic cycle length, BCL), each with initial
conditions 
\begin{gather}
    \label{stimvolt2}
  \u(k\B)=\u_\text{stim},~~~~\w(k\B)=\w\big((k-1)\B\big)=\wB.
\end{gather}
\end{subequations}
The statement that 
$\w(k\B)=\w\big((k-1)\B\big)=\wB$
for some $k>m$ 
will be justified in
subsection \ref{sec:periodicstimulation}, below.

Equations \eqref{McKean} were proposed by \cite{McKean1970} as a model
of the action potential of spiking neurons. They are qualitatively
equivalent to the FitzHugh-Nagumo
equations \citep{FitzHugh1961,Nagumo1962}, the latter being in turn a 
reduction of 
the pioneering \cite{Hodgkin1952} model of the action potentials in
the squid giant axon. These equations been extensively studied in the literature
e.g.~\citep{Rinzel1973,Wang1988a,Wang1988,Tonnelier2003,Bezekci2015}
and below we recall, briefly and informally, some elements of their geometric singular
perturbation analysis with the aim of determining (a) the parameter range
where the model is excitable, and (b) an asymptotic expression for the
APD, both necessary for modelling of experimental results in subsequent sections.

\subsection{Phase portrait}
\begin{figure*}[t]
\begin{overpic}[width=0.482\textwidth,tics=10]{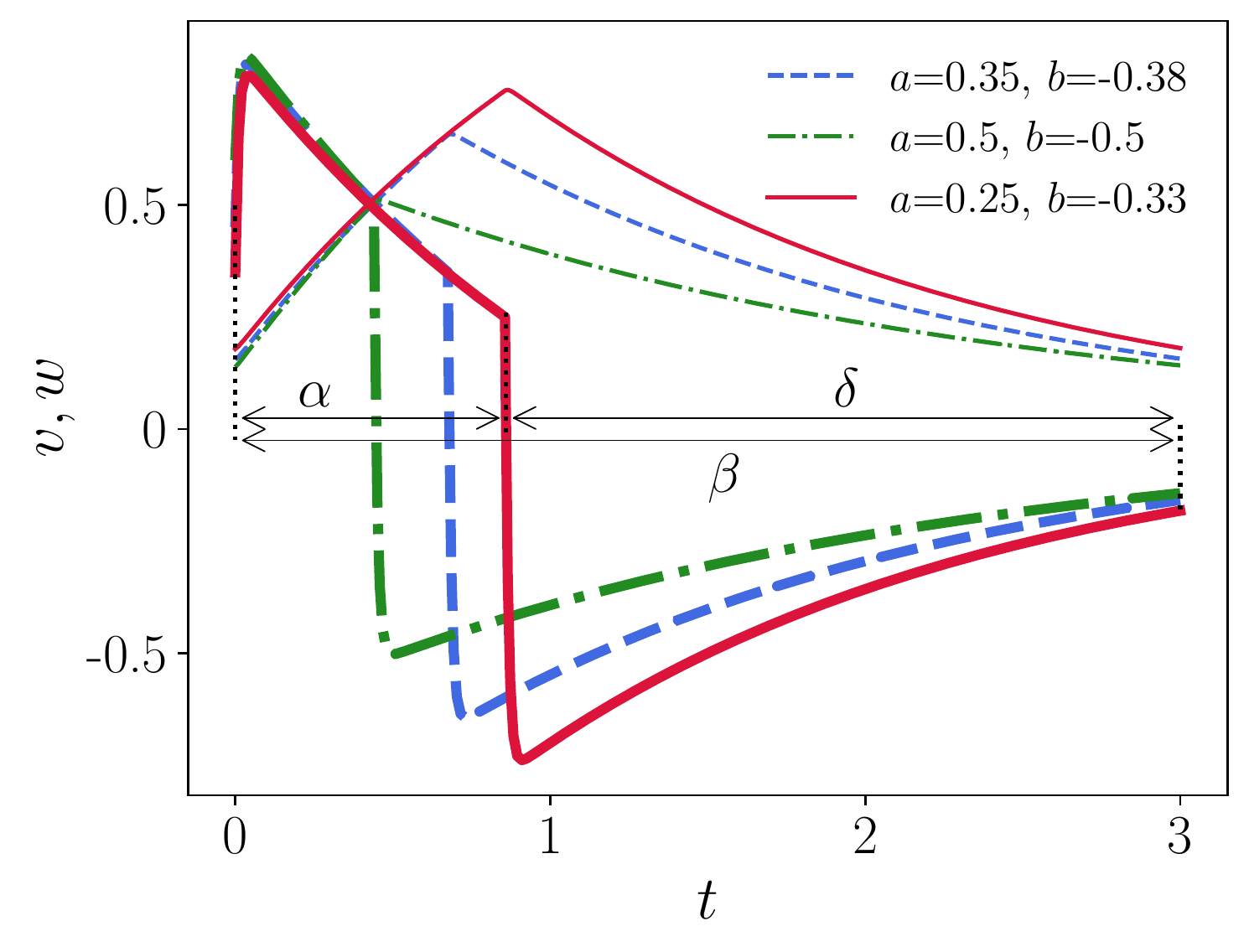}
\put (2,71) {{(a)}}
\end{overpic}
\hfill
\begin{overpic}[width=0.49\textwidth,tics=10]{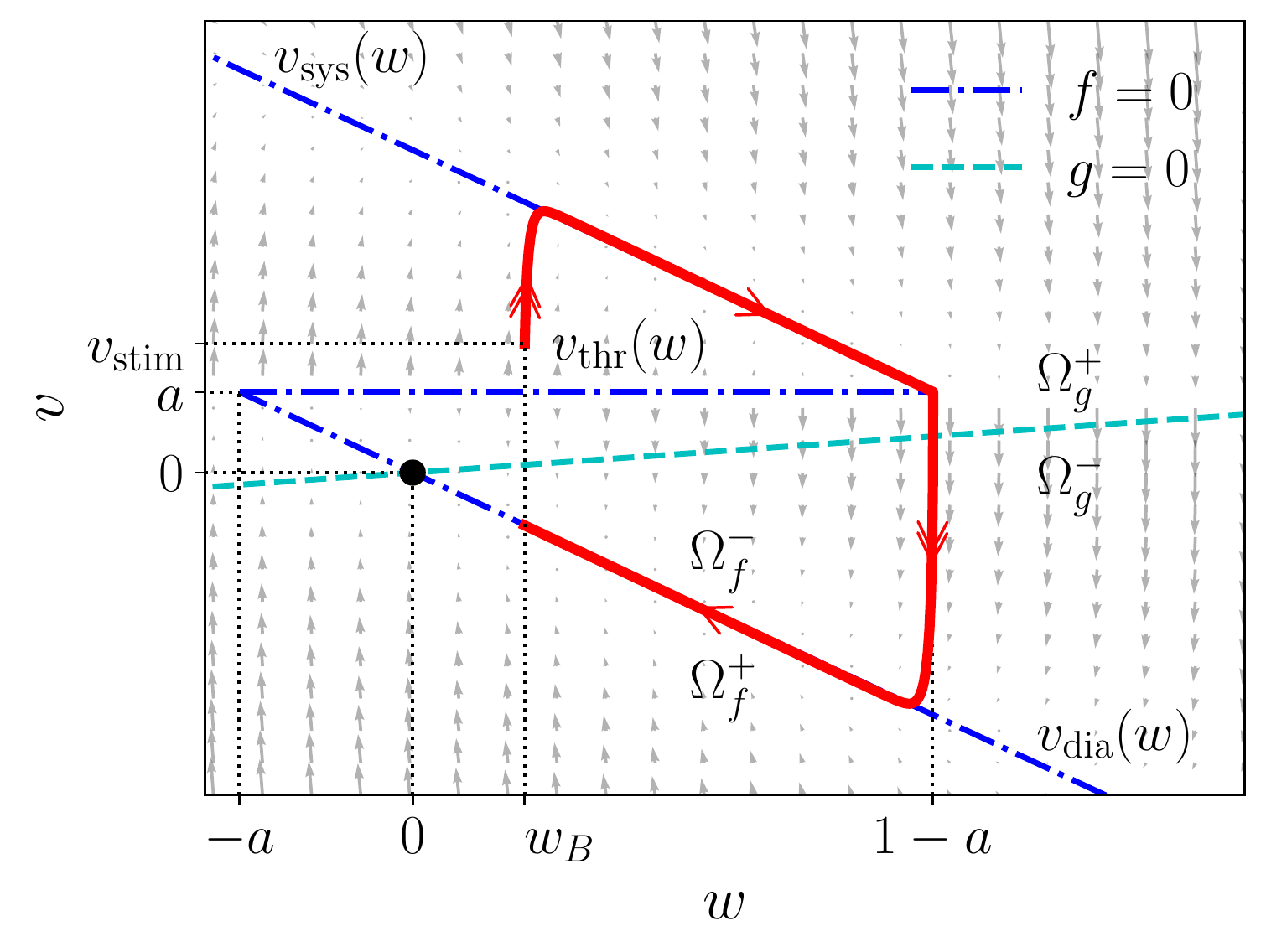}
\put (2,70) {{(b)}}
\end{overpic}
\caption{
(a) Three examples of action potential solutions to the McKean
equations \eqref{McKean}. The ``voltage''
$\u(\t)$ (thick curves)
and the effective ``gating variable'' $\w(\t)$ (thin curves) are
shown as functions of time for parameter values $\eps=0.01$, $\B=3$
and randomly selected combinations of $\a$ and
$\c$ as shown in the legend. The action potential duration $\A$, diastolic 
interval $\D$, and basic cycle length $\B$ are annotated on one of the
AP curves.
(b) Phase portrait and vector field of the McKean
equations \eqref{McKean} and associated notation.
The nullclines $\fu=0$ and $\g=0$ are shown as a dash-dotted blue
line and a dashed turquoise line, respectively. The single
attracting global equilibrium $(0,0)$ is marked with a black dot marker. A typical trajectory
is shown in a solid red line where a double arrow indicates a fast
piece and a single arrow indicates a slow piece of the trajectory. Parameter values
used are $\a = 0.25$, $\c = 0.3$, $\eps = 0.01$ and correspond to 
excitable dynamics.
}
\label{fig01}
\end{figure*}

The solutions to the \cite{McKean1970} model have a
generic action potential waveform as shown in Figure \ref{fig01}(a),
c.f.~figure 1 and 2 of \citep{Lachaud2022} for comparison to waveforms measured
in experiments. To understand these solutions we consider the singular asymptotic limit $\eps\to0^+$ in
which equations \eqref{McKean} reduce to a fast-time subsystem
 \begin{gather}
 \label{McKean.fast}
   \df{\u}{\T} = \fu(\u,\w), ~~~~~
   \df{\w}{\T} =  0,
 \end{gather}
when written in terms of the ``fast'' time variable $\T := \eps^{-1}\t$, and to a
slow-time subsystem
 \begin{gather}
 \label{McKean.slow}
   0 = \fu(\u,\w), ~~~~~
   \df{\v}{\t} =  \fv(\u,\w),
 \end{gather}
\looseness=-1
when written in terms of the original ``slow'' time variable $\t = O(1)$.
The nullcline $\f(\u,\w)=0$ plays a special role in both
systems \eqref{McKean.fast} and \eqref{McKean.slow}, and it is known as
the ``critical set'' or the ``reduced slow set'' of
\eqref{McKean}, because it is in one-to-one correspondence to fixed points of the fast
subsystem \eqref{McKean.fast}, and 
because the trajectories of the slow subsystem \eqref{McKean.slow} are
constrained to follow it.
The critical set, $\fu(\u,\v)=0$, is a piece-wise
linear caricature of a cubic function of $\u$ as illustrated in Figure \ref{fig01}(b).
Specifically, it has local minima and maxima at the points
$\Mmin=(\a,-\a)$ and $\Mmax=(\a,1-\a)$, respectively, and roots 
\begin{align}
&\begin{cases}
\uplus=1-\w & \text{for}~~ \w \in(-\infty,-\a],\\
\umins=-\w,~~ \uth=\a,~~\uplus=1-\w
& \text{for}~~ \w \in[-\a,1-\a],\\
\umins=-\w & \text{for} ~~ \w \in[1-\a,\infty).
\end{cases}
\end{align}
\looseness=-1
Since $\fu(\u,\w) <0$ in the region
$\Omega_f^- :=\big\{(\u,\w)\in \mathbb{R}^2 : \w>\bar\w ~\text{where}~ \fu(\u,\bar{\w})=0\big\}$,
and $\fu(\u,\w) > 0$ in the region 
$\Omega_f^+ :=\big\{(\u,\w)\in \mathbb{R}^2 : \w<\bar\w
~\text{where}~ \fu(\u,\bar{\w})=0\big\}$, the branches
$\uplus(\w)$ and $\umins(\w)$, called ``systolic'' and ``diastolic''
respectively, consist of stable attracting fixed points and the
``threshold'' branch $\uth(\w)$ consists of unstable repelling fixed
points of the fast-subsystem \eqref{McKean.fast}.
The second nullcline $\fy(\u,\v)=0$ is a straight line which
partitions the phase plane in two regions
$\Omega_g^- :=\big\{(\u,\w)\in \mathbb{R}^2 : \w>\bar\w ~\text{where}~ \fy(\u,\bar{\w})=0\big\}$
where  $\fy(\u,\w) <0$, and
$\Omega_g^+ :=\big\{(\u,\w)\in \mathbb{R}^2 : \w<\bar\w  ~\text{where}~ \fy(\u,\bar{\w})=0\big\}$
where $\fy(\u,\w) > 0$
and thus determines the direction of
the slow flow of \eqref{McKean.fast} along the critical set.
These facts are illustrated by the vector field shown in Figure \ref{fig01}(b).
Thus, in the singular approximations given by \eqref{McKean.fast}
and \eqref{McKean.slow} a typical trajectory
of the McKean model \eqref{McKean} consists of fast jumps to one of the attracting
branches of the critical set followed by slow motions to the end
the attracting region or until a globally stable fixed point is
reached as illustrated in Figure \ref{fig01}(b).
Global fixed points occur at the intersection of the two nullclines and
apart from degenerate cases there exists either one single or three
distinct fixed points given by 
\begin{align}
\label{critical.branches}
(\u_\ast,\w_\ast) \in
\begin{cases}
\big\{(0,0)\big\} & \text{if} ~~ \c < \a/(1-\a),\\
\big\{(0,0),~ (\a, \a/\c), ~\big(\c/(1+\c),1/(1+\c)\big)\big\} & \text{if} ~~\c \ge \a/(1-\a).
\end{cases}
\end{align}

\subsection{Asymptotic approximation of excitability}
\begin{figure*}[t]
\raisebox{0mm}{\begin{overpic}[width=0.5\textwidth,tics=10]{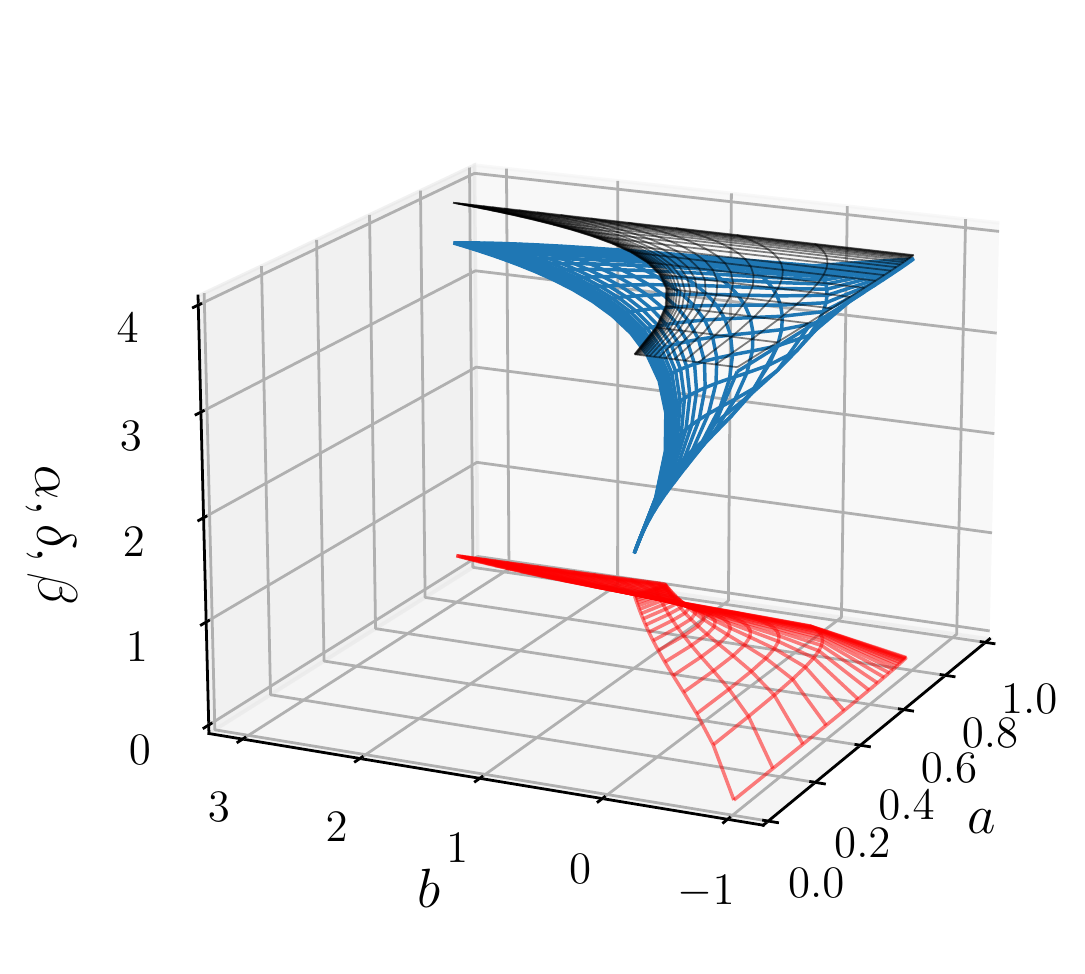}
\put (3,70) {{(a)}}
\end{overpic}}\hfill
\raisebox{0mm}{\begin{overpic}[width=0.5\textwidth,tics=10]{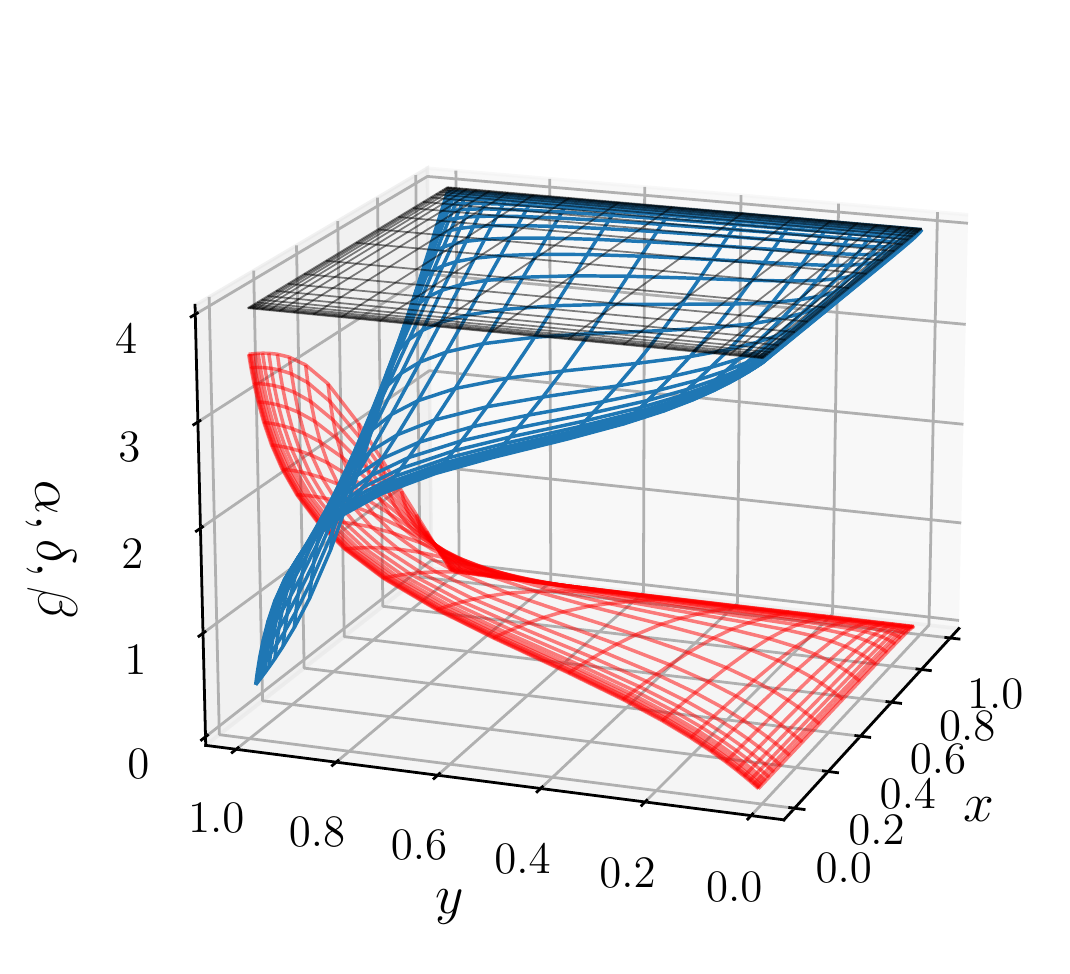}
\put (3,70) {{(b)}}
\end{overpic}}
\caption{
Asymptotic expressions \eqref{APD.B}  and \eqref{APD.D} for the action
potential duration $\A$ (red
wire-frame) and diastolic interval $\D$ (blue
wire-frame), respectively, (a) as functions of the McKean 
model parameters $\a$ and $\c$ and (b) as functions of the
``rectangular'' variables $x$ and $y$ defined in equation \eqref{chvars}.
The value of the basic cycle length is $\B=4$ and the surfaces
$\B=\A+\D$ (black wire-frame) are plotted as a test and to
illustrate the parameter space $\Omega_\text{ex}$ for excitable
dynamics given by \eqref{excitability.region}.}
\label{FIG020}
\end{figure*}

\looseness=-1
Biological excitability is usually described as a reaction of a system
to an external stimulus that invokes a sufficiently large finite
response before return to a unique global equilibrium. It follows that in McKean's 
model \eqref{McKean}, excitability 
corresponds to the case of 
a single stable attractor located 
between the fold  points $\Mmin$ and $\Mmax$
on the diastolic branch $\uplus(\w)$ 
and hence to a parameter space given by
\begin{align}
\label{excitability.region}
\Omega_\text{ex} = \Big\{ (\a,\c) \in \mathbb{R}^2 : 
~~~ \c> -1  ~~\cap ~~
\c< \a/(1-\a) ~~\cap~~
\a>0 ~~\cap~~
\a<1\Big\}.
\end{align}
Region $\Omega_\text{ex}$ is visualised in Figure \ref{FIG020}(a).
Indeed, in this case starting from the initial
conditions \eqref{stimvolt2}, a trajectory performs a fast jump of
infinitesimally short duration $O(\eps)$ from point
$(\u_\text{stim}, \wB)$ to point
$\big(\uplus(\wB), \wB\big)$ governed by
the fast subsystem \eqref{McKean.fast}.  Next, it follow the
systolic branch from the latter point to point $\Mmax$  for a period of
duration found by integrating the slow subsystem \eqref{McKean.slow}, 
\begin{align}
\label{APD.wB}
\A &:= \int_{\wB}^{1-\a} \d \t 
= \int_{\wB}^{1-\a} \frac{\d\w}{\fy\big(\uplus(\w),\w\big)}
=\frac{1}{1+\c}\log\left(\frac{1-(1+\c)\;\wB}{1-(1+\c)(1 - \a)}\right).
\end{align}
At the fold point $\Mmax$ the systolic branch of the critical set
$\uplus$ terminates and switches to the repelling threshold branch $\uth$ so
the trajectory makes another infinitesimally short fast jump over to
point $\big(\umins(1 -\a),1-\a\big)$. Finally, the trajectory follows the
attracting diastolic branch $\umins(\w)$ towards the single global stable
fixed point $(0,0)$ for period of duration found by integrating the slow subsystem \eqref{McKean.slow},
\begin{align}
\label{DI.wB}
\D &:= \int_{1-\a}^{\wB} \d\t
= \int_{1-\a}^{\wB} \frac{\d\w}{\fy\big(\umins(\w),\w\big)}
= \frac{1}{1+\c}\log{\left(\frac{1 - \a }{\wB}\right)},
\end{align}
when it receives its next excitation stimulus at $(\u_\text{stim},\wB)$ and repeats
these motions as illustrated in Figure \ref{fig01}(b). 
Our presentation has been rather informal, so we remark that the
asymptotic analysis of fast-slow systems, such as the McKean
model \eqref{McKean}, has a rigorous foundation grounded in classical theorems due to
\cite{Tikhonov-1952,Pontryagin-1957,Fenichel-1979} and a good
exposition with an extensive list of references may be found in \citep{Kuehn2015}.
\begin{figure*}[t]
\raisebox{0mm}{\begin{overpic}[width=0.5\textwidth,tics=10]{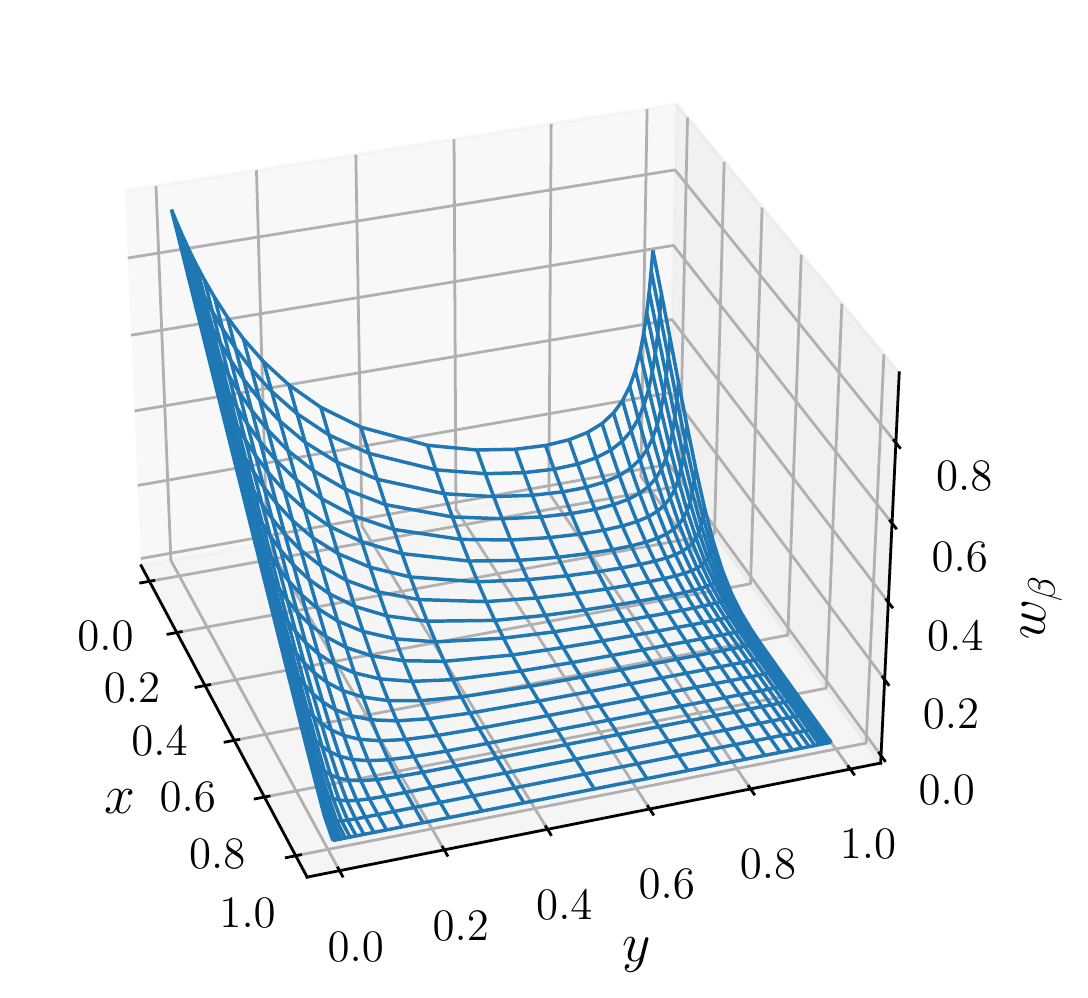}
\put (3,75) {{(a)}}
\end{overpic}}\hfill
\raisebox{6mm}{\begin{overpic}[width=0.45\textwidth,tics=10]{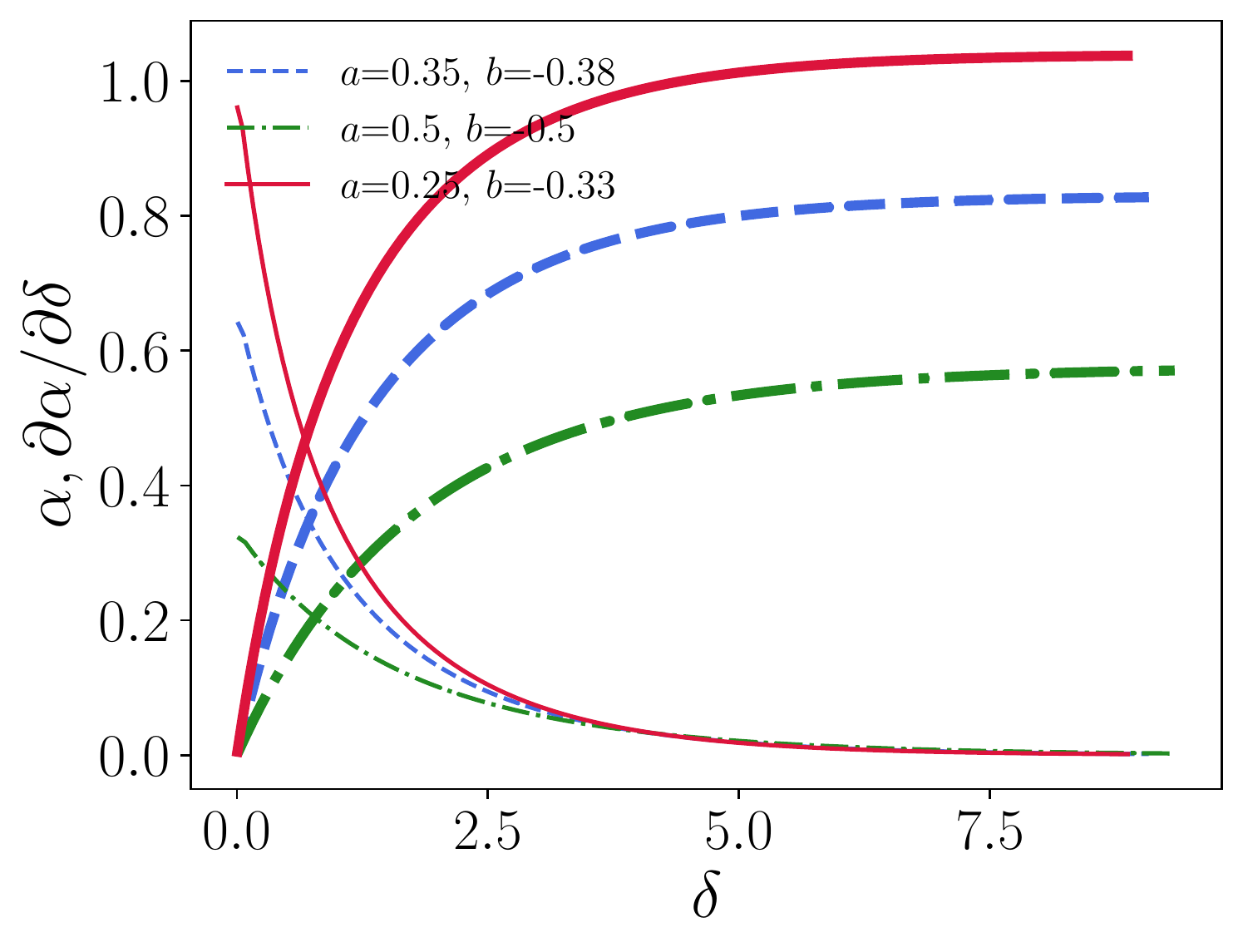}
\put (0,75) {{(b)}}
\end{overpic}}
\caption{
(a) The initial condition $\wB$ required for conformance to
stimulation with fixed period $\B=4$ given by equation \eqref{wB} as a function of the 
``rectangular'' variables $x$ and $y$ defined in \eqref{chvars}.
(b) Examples of asymptotic \rev{restitution
curves $\A$ as a function of $\D$ (thick lines) and their derivatives $\partial_\D \A$}
(thin lines) for selected McKean  parameter values given in the legend.}
\label{FIG030}
\end{figure*}

\subsection{APD restitution under periodic stimulation}
\label{sec:periodicstimulation}
\looseness=-1
So far, we have not  discussed the parameter $\wB$ introduced
in the initial conditions \eqref{stimvolt2}. In fact, this is not an
independent parameter, rather it is determined by the prescribed
basic cycle length $\B$. Conditions for uniqueness and existence of solutions
to an initial-value problem require that for the solutions of
equations \eqref{McKean} 
to be identical up to a time shift (i.e. periodic APs) on two consecutive time intervals
$\t\in\big(k\B,(k+1)\B\big]$, $k\in \mathbb{N}$,
they must start from identical
initial conditions. This implies that a periodic 
train of action potentials exists only when the sequence
$\big\{\w(k\B), ~k=0,1,\dots\}$ converges to a unique value $\wB$,
possibly after a transient period of a finite number of 
stimuli $k>m$. There may exist parameter values including that of
$\B$, for which the sequence does not converge but generates more complex
behaviour such as alternans.
\rev{Alternans, (more precisely, APD alternans, as other types also exist) is a
regime of myocyte response to periodic stimulation where action
potentials alternate in duration between long and short even though
the pacing cycle length remains constant, see e.g.~\citep{Qu2014}.} 
However, investigating these
possibilities is beyond the scope of the current analysis, and here we
assume a strictly periodic response to the external
stimulation. Under this assumption and proceeding to employ asymptotic
approximation, we neglect the duration of any fast jumps as being of order
$O(\eps)$ and identify expression \eqref{APD.wB} as the duration of
the action potential, \APD{}, and expression \eqref{DI.wB} as its diastolic
interval, DI. Thus, we require, that the sum of the APD and the DI equals
the BCL,
\begin{align}
\label{BB}
\B=\A+\D 
&=
\frac{1}{1+\c}\log\left(\frac{\big(1-(1+\c)\wB\big)(1 - \a)}{\big(1-(1+\c)(1 - \a)\big)\wB}\right).
\end{align}
Solving this  algebraic equation yields the value of $\wB$
necessary to establish a periodic train of APs,
\begin{align}
\label{wB}
\wB = \Big((1+\c)+\exp\big((1+\c)\B\big)\big(1-(1+\c)(1-\a)\big)/(1-\a)\Big)^{-1}.
\end{align}
Substituting \eqref{wB} into \eqref{APD.wB}, we arrive at an explicit closed
form \rev{expressions for the \APD{} and the DI as  functions} of the intrinsic model
parameters $\a$ and $\c$ and the protocol dependent basic cycle length
parameter $\B$,
\begin{subequations}
\begin{align}
\label{APD.B}
\A(\a,\c,\B) &= 
\frac{1}{\cnew}\log\left(\frac{\exp\big(\cnew\B\big)}{\anew\cnew+\big(1-\anew\cnew\big)\exp\big(\cnew\B\big)}\right),
~~~~ \anew:= 1-\a, ~~\cnew:=1+\c,\\
\label{APD.D}
\revm{\D(\a,\c,\B)} &= \revm{\B - \A(\a,\c,\B),}
\end{align}
\end{subequations}
with $\anew$ and $\cnew$ introduced for brevity only.
\rev{These expressions} along with equation \eqref{BB} are
illustrated in Figure \ref{FIG020} as surfaces over the parameter
region of excitable dynamics, $\Omega_\text{ex}$. The outline of
$\Omega_\text{ex}$ seen in Figure \ref{FIG020}(a) and illustrates the
fact that one of its boundaries, $\a=1$, is an asymptote to another,
$\c=\a/(1 -\a)$. This makes visualisation of results in the $(\a,\c)$
parameter plane difficult and we introduce a change of variables
\begin{align}
\label{chvars}& \a = x, ~~~~~ \c = y/(1-x)-1, 
~~~~~ x\in(0,1), ~~~~~  y\in (0,1),
\end{align}
that maps $\Omega_\text{ex}$ into the rectangular domain $(x,y) \in
(0,1) \times (0,1)$. The result is
illustrated in Figure \ref{FIG020}(b) which is identical to
Figure \ref{FIG020}(a) but uses $x$ and $y$ as independent variables.
We remark that for fixed values of the parameters $\a$ and $\c$,
\rev{the relation between \APD{} and DI} is known as \rev{the
restitution curve} and is widely used in physiological 
experiments to infer the stability of periodic AP trains from the
condition that alternans occur for \rev{values of $\D$} for which
the slope of the restitution curve is greater than unity,
\rev{i.e. $|\partial_\D \A| > 1$, see e.g.~\citep{Qu2014} and references within.}
\rev{An asymptotic restitution curve for the McKean model can be
obtained in parametric form from expressions \eqref{APD.B} and \eqref{APD.D} 
as $\big(\D(\a,\c,\B),\A(\a,\c,\B)\big)$ with the
basic cycle length $\B$ taking the role of the parameter along the curve.}
Restitution curves and their gradients are plotted in
Figure \ref{FIG030}(b) for several randomly selected combinations \rev{of
parameter values} $\a$
and $\c$  and show the generic shape known from experimental measurements with the
slope indicating that instabilities do not occur at these parameter values.

\looseness=-1
Figure \ref{FIG030}(a) shows the value of $\wB$ found from
expression \eqref{wB} as a function of the parameters $\a$ and $\c$
at a fixed basic cycle length $\beta$.  The figure illustrates a fact
that is perhaps insufficiently appreciated.
\rev{
Each model in a population of uncoupled non-identical McKean models
is characterised by different values of $\a$ and $\c$. Thus, each model
requires a different value of $\wB$  to produce a stable periodic 
response to a common pacing sequence. Since $\wB$ is the
value of the effective gating variable at the moment of stimulation,
and channel blocking drugs seek to alter the proportion of open
channels, individual myocytes will be affected to a different extent
by a  drug dosage common to all cells.
}
  
We remark that several other variants of the McKean kinetics have been
proposed e.g.~\citep{McKean1970,Barkley1991,Fall2004} and along with
the FitzHugh-Nagumo model these would yield qualitatively equivalent
results, even though not in the convenient closed form found here.
Finally and most importantly, we note that the McKean
equations \eqref{McKean} are an appropriate phenomenological model
of the rabbit ventricular APD restitution because the archetypal
asymptotic structure of realistic cardiac AP models
includes a conventional Tikhonov slow-time subsystem of McKean type
even though it is essentially non-Tikhonov overall as demonstrated by \cite{Biktashev2008}. 

\section{Application of McKean asymptotics to ion-channel block experiments}
\label{application}

\subsection{Model of drug action}
\looseness=-1
We now proceed to apply the asymptotic results obtained in the
preceding section to the ion-channel block experiments
of \cite{Lachaud2022}. In these experiments over 500 myocytes were
isolated from rabbit left ventricular walls and significant
cell-to-cell variability in APD was established for the first time. Although cells were
extracted from different apical/basal and endo/mid/epicardial sub-regions
of 12 different animal hearts, we will treat them here as
a single large and diverse myocyte population. The myocytes were 
loaded with a voltage-sensitive dye and subjected to a periodic
excitation stimulus with basic cycle length $500$ ms
(i.e.~experimental stimulation frequency of 2 Hz).
In a second stage of the experiment, 
1 $\mu$M of the drug nifedipine was applied to cells to
examine their response to ion channel blocking drugs. 
Fluorescence signals were recorded from all cells both before
and after drug application and selected AP waveform biomarkers, principally
the action potential duration \APDn, were 
measured for every cell able to follow the stimulation protocol for
more than 4 min. Detailed description of the
experimental methods, quality control protocols along with descriptive
statistics of sub-populations, estimates of experimental error, and
experimental datasets can be found in \citep{Lachaud2022}.

We assume that myocytes can be described mathematically by the 
McKean equations \eqref{McKean}. Let 
\begin{align}
\De(\Bexp,\beexp)=\Big\{ \Big(\Aexp_i(\Bexp), ~\big(\Aexp_i(\Bexp)+\dAexp_i(\Bexp,\beexp)\big)\Big), ~~~~
i=1,\dots, N\Big\}
\end{align}
be a set of experimental data points consisting of pairs of 
action potential duration values measured before and after drug application, $\Aexp_i$
and $(\Aexp_i+\dAexp_i)$ respectively, for each cell 
$i=1,\dots, N$, at a fixed experimental basic 
cycle length $\Bexp$ and drug concentration $\beexp$. Here,
$\Bexp=500$ ms, $\beexp = 1$ {$\mu$M nif.},~and $N=496$. Let 
\begin{align}
\Pe(\B,\be)=\Big\{ \big(\a_i, \c_i\big), ~~~~ i=1,\dots, N\Big\}
\end{align}
be a set of corresponding McKean model parameter pairs $\a_i$ and
$\c_i$ to be determined for each cell $i$. We relate the action potential
duration values before and after drug application to the McKean
model parameters by the following
set of $2N$ non-linear equations
\begin{subequations}
\label{experiment}
\begin{align}
\label{experiment.before}
\Aexp_i/\Bexp&=\A(\a_i,\c_i,\B)/\B,\\
\label{experiment.after}               
(\Aexp_i+\dAexp_i)/\Bexp&=\A(\be \a_i,\c_i,\B)/\B, & i=1,\dots, N,
\end{align}
\end{subequations}
where $\A(\cdot)$ is the asymptotic expression \eqref{APD.B} for
the action potential duration, $\B$ is the value
of the basic cycle length, and $\be>0$ is a parameter representing drug
action in the McKean model, respectively.
Since myocytes are uncoupled in the experiment, these $2N$ equations
decouple to a set of $N$ independent pairs of equations one per each
cell, and for brevity, subscripts $i$ will be used 
only to denote specific cell values and omitted when discussion
is valid for any arbitrary cell.
Drug action is encoded in 
equations \eqref{experiment.after} by introducing the
multiplicative factor $\be$ in front of the parameter $\a$.
Indeed, even though the McKean model has no relation to realistic
electrophysiological processes and structures in the cardiomyocytes,
the parameter $\a$ plays a role similar to that of a channel current
conductance: it is the only parameter on the right-hand side of the
McKean ``voltage'' equation \eqref{McKean-Eq1}, which in
turn represents the total sum of all ionic currents in physiologically-realistic models.
The second McKean parameter $\c$ controls the rate of change
of the effective gating variable $\w$ via equation \eqref{McKean-Eq2}, and so must be
interpreted as an effective kinetic parameter.
In the following, we will further restrict the attention to the case of
$\be>1$ to compare with available nifedipine data. Note, that $\be>1$
does not mean that ion channels are enhanced but only that the
$\u$-nullcline is translated up the $\u$-axis.
Finally, the factor $\Bexp/\B$ is introduced to convert experimental and
model \APD{} values, $\Aexp$ and $\A$, which are measured in different
units, to a common non-dimensional form by scaling these with the
experimental and model cycle lenghts $\Bexp$ and $\B$, respectively.

\rev{Dofetilide data is also reported in \citep{Lachaud2022} but will
not be modelled  here. Dofetilide has a more complex effect: it
prolongs the \APD{} of some cells while simultaneously shortening the APD
of other cells, see Figure 3B(iii) of \citep{Lachaud2022}.
\cite{Crumb2016} and \cite{Li2017} suggest that dofetilide is
a multi-channel blocker that acts on several distinct ion channels. So
considering the case $\gamma < 1$, that only prolongs \APD{}, is
insufficient to capture the 
effect of this drug. To model the effect of dofetilide two distinct 
``effective conductance'' parameters like $\a$ are required, one of
which shortens and the second of which prolongs the \APD{}, so that 
in combination they reproduce the measured response to dofetilide. 
However, only one such parameter, $\a$, is present in McKean's model. 
The second McKean parameter, $\c$, is an effective kinetic
parameter and cannot be used for this purpose.}

\rev{
The multiplicative model $\bar{\a}=\be\a$, where $\a$ and $\bar{\a}$
are the values of the McKean parameter $\a$ before and after drug
administration, arises as a linearisation of the general relation $\bar{\a}=\bar{\a}(\beexp)$
of the parameter $\a$ as a function of the drug concentration
$\beexp$. Indeed, for sufficiently small drug concentration values
$$
\bar{\a}=\bar{\a}(\beexp) \approx \big(1+k\beexp\big)\a, ~~~~~~~
\a=\bar{\a}(0),
~~~k=\frac{\partial \bar{\a}}{\partial \beexp}(0),
$$
so that the multiplication factor $\be$ depends linearly on the drug
concentration 
$$
\be:=\big(1+k\beexp\big).
$$
The constant  $k$ can be eliminated using a suitable calibration
$\be^\ast=\big(1+k\beexp^\ast\big)$,
so that an explicit relation between the drug
concentration $\beexp$ and the drug action parameter $\be$ can be
found,
$$
\beexp=\beexp^\ast\frac{\be-1}{\be^\ast-1}.
$$
This result is used in equation \eq{eq:scaling} while 
calibration values $\be^\ast$ and $\beexp^\ast$ are discussed in
section \ref{calib} below.
Our multiplicative model bears similarity to the so called ``conductance-block''
model of drug action widely used with realistic cardiac ionic current
models. In realistic models an ion channel current takes the form
$$I_j = g^0_j O (V - E_j),$$
where, $g^0_j$ is the maximal conductance of a population of fully
open channels of type $j$, $O$ is its open probability,
and $E_j$ is the reversal potential for the species of ion which flows
through these channels. To account for drug action the maximal
conductance $g_j$ is  multiplied by the factor 
$$
\be_\text{ion}=K_d/(K_d+\Gamma),  
$$
representing the percentage $C/C_0$ of channels $C$ remaining unbound
in the presence of the drug. The latter expression is
obtained from a steady-state approximation of the law of mass
action for the reaction
$$C+\beexp  \xrightleftharpoons[k_+]{k_-} B,$$
where $k_\pm$ are reaction rates, $C$ and $B$ are open and bound
channels with $C+B=C_0$ being the entire channel population,
e.g.~\citep{Keener2009}. The 
equilibrium constant $K_d:=k_-/k_+$ measures the potency of the drug and
is often approximated by the half-maximal inhibitory concentration
(IC50) measured in experiments. The `conductance-block'' formulation
is widely used in numerical studies \citep{Brennan2009,Mirams2011},
but it cannot be applied directly here as the McKean parameter $a$ is not
an ion channel conductance even though it effectively plays the role
of one.}

\subsection{Domain and existence of solutions}
\label{existence}
\begin{figure*}[t]
\raisebox{0mm}{\begin{overpic}[width=0.495\textwidth,tics=10]{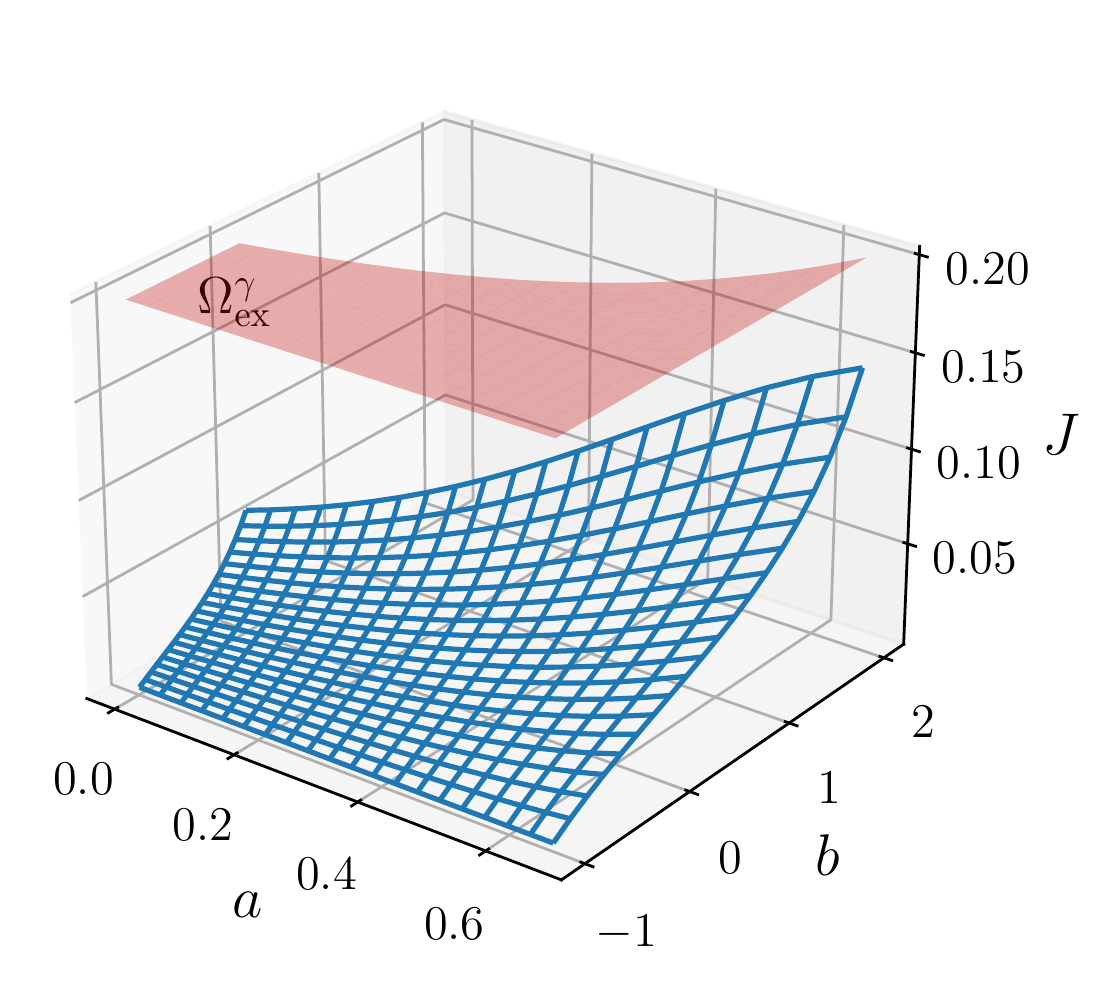}
\put (1,73) {{(a)}}
\end{overpic}}\hfill
\raisebox{2mm}{\begin{overpic}[width=0.495\textwidth,tics=10]{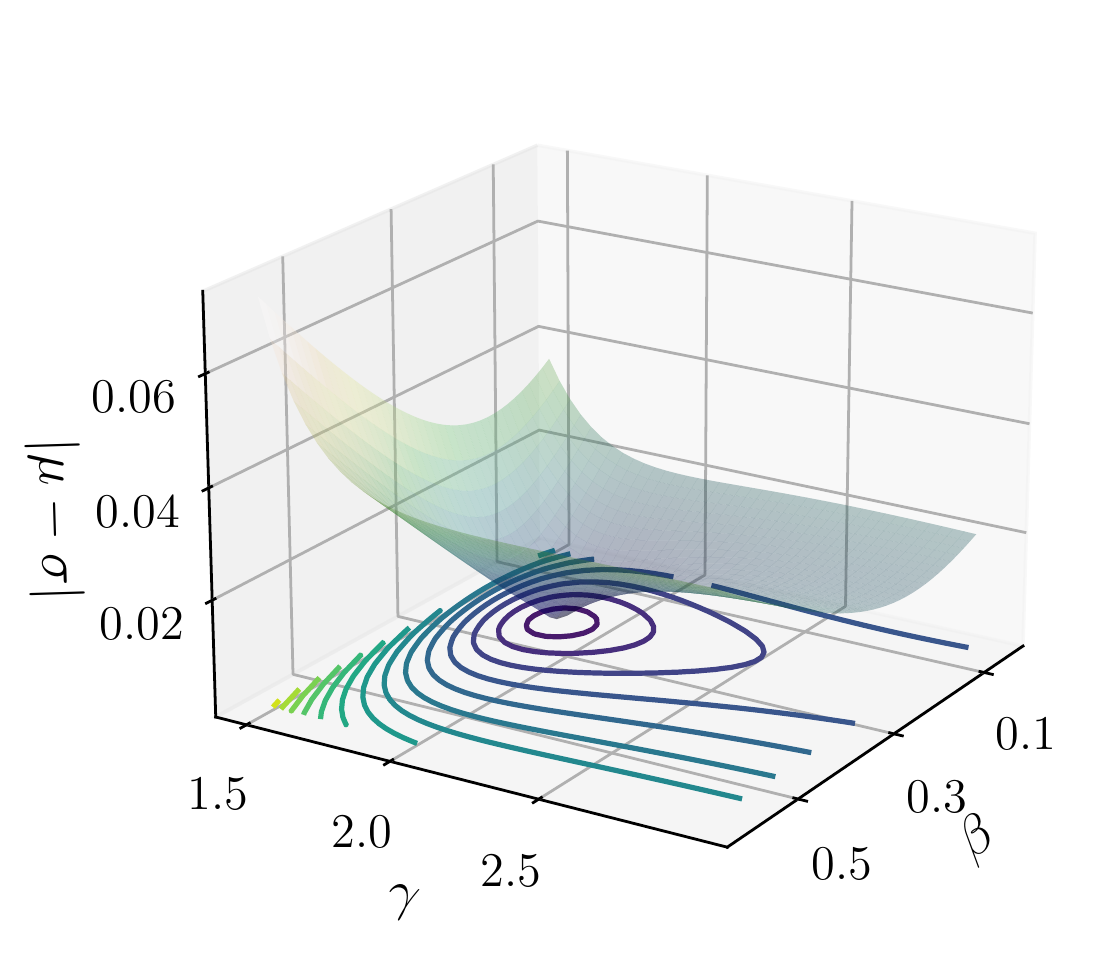}
\put (1,70) {{(b)}}
\end{overpic}}
\caption{
\looseness=-1
(a) The Jacobian determinant
$J= \det \big(\partial(F_1,F_2)/\partial(\a,\c)\big)$ as a function of
parameters $\a$ and $\c$ for $\B=0.3$ and $\be=1.5$ is shown by a
blue wire frame, with the grid lines being iso-lines of the
rectangular coordinates $x$ and $y$ defined in \eqref{chvars}. The pink transparent region at the top of the axes
box is the domain of excitability ${\Omega}^\gamma_\text{ex}$.
(b) Convexity and global minimum of the distance
$|\vec{\mu}-\vec{\sigma}|$ between the centre of mass of the set of parameter values $\Pe$ and that of the
planar region $\Omega^\gamma_\text{ex}$, 
see \eqref{calibrateBandBe}. The values of $\Pe$ are determined
as solutions to \eqref{experiment} for the set of experimental
measurements $\De$ of \cite{Lachaud2022}.}
\label{FIG040}
\end{figure*}

\begin{figure*}[t]
\raisebox{0mm}{\begin{overpic}[width=0.497\textwidth,tics=10]{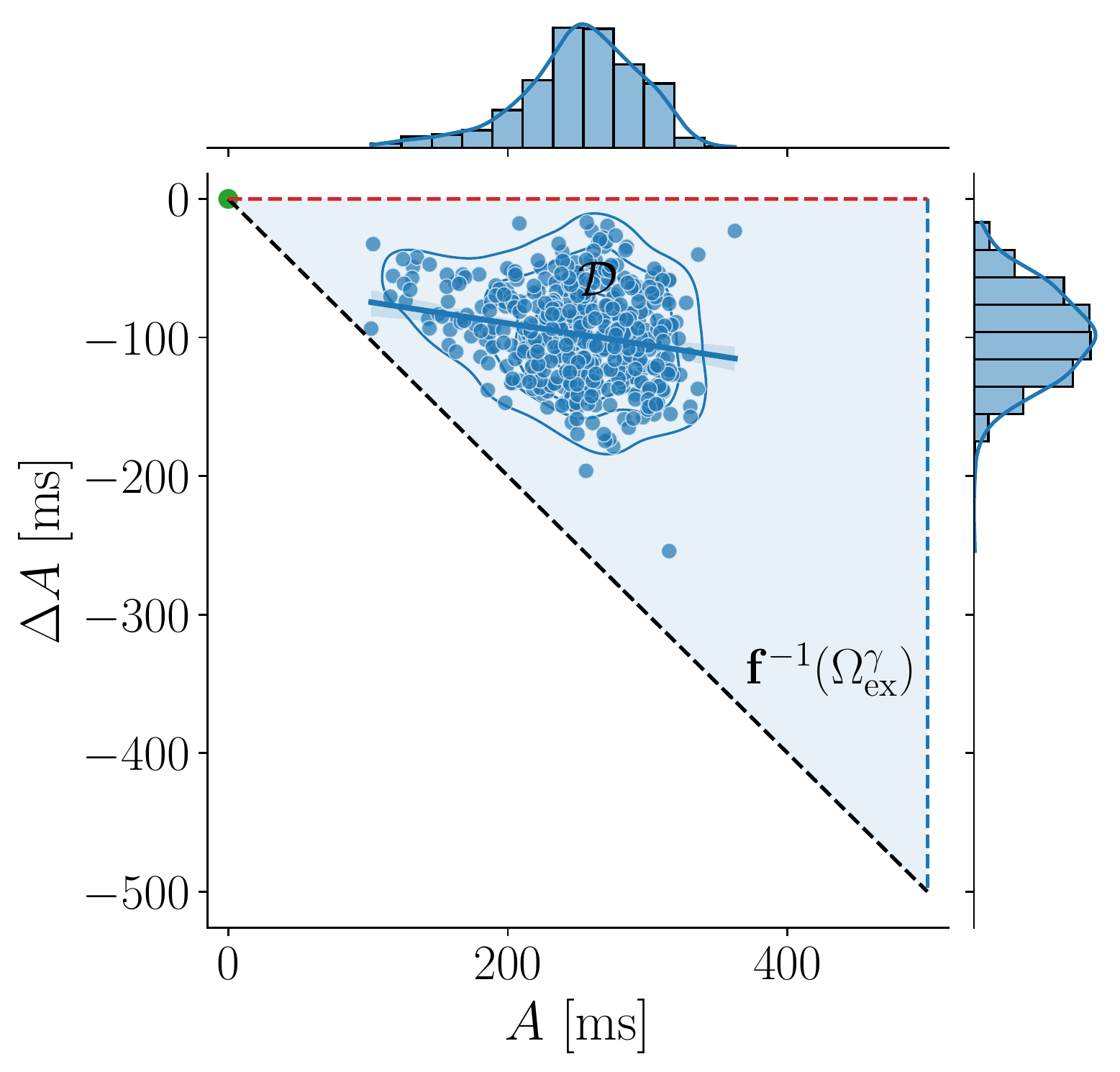}
\put (1,84) {{(a)}}
\end{overpic}}\hfill
\raisebox{0mm}{\begin{overpic}[width=0.498\textwidth,height=0.479\textwidth,tics=10]{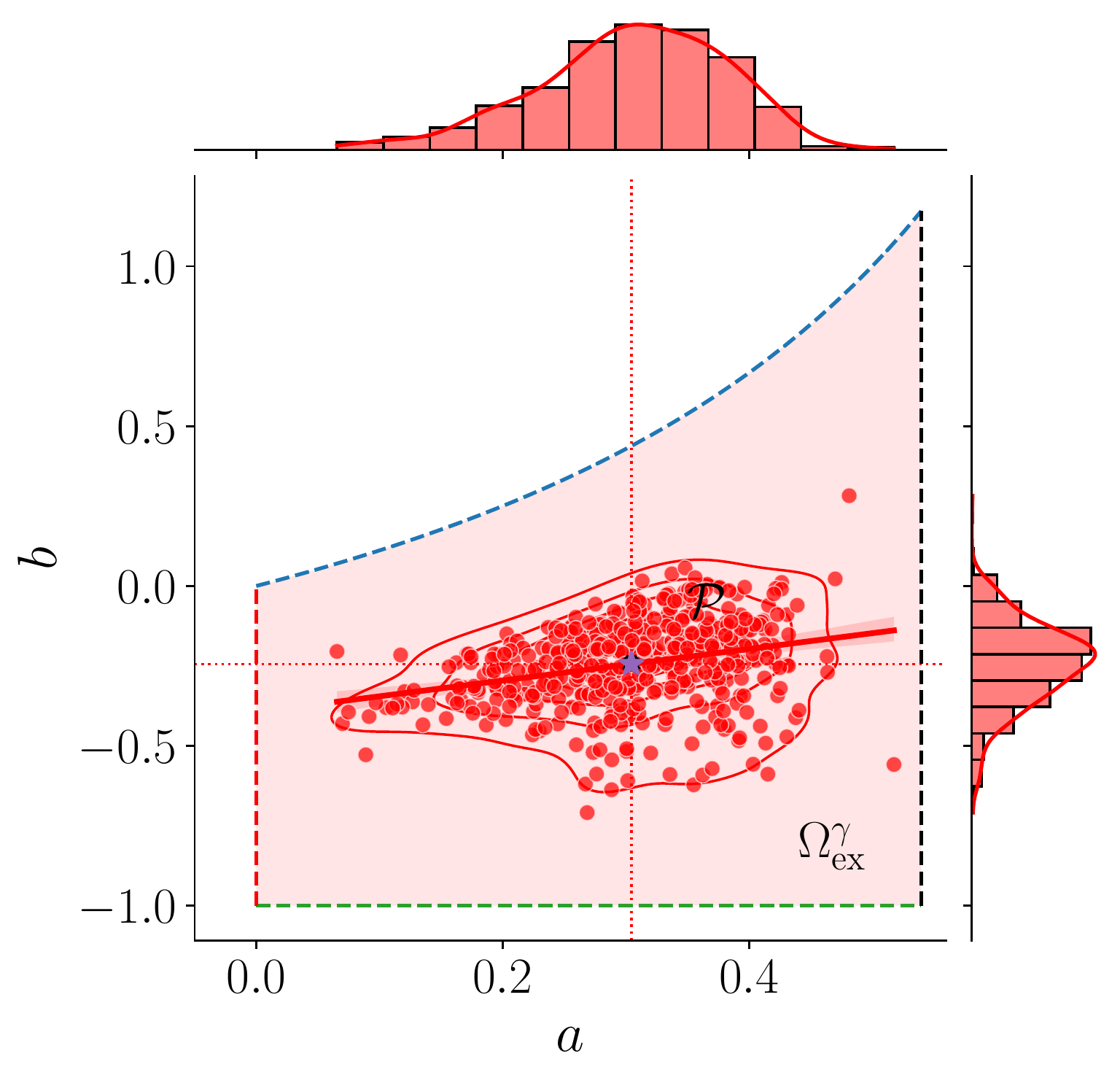}
\put (1,84) {{(b)}}
\end{overpic}}
\caption{
(a) Scatter plot of the set $\De(\Bexp,\beexp)$ of 496 experimental
measurements at $\Bexp=500$ ms and $\beexp=1$ $\mu$M nifedipine, data due
to \citep{Lachaud2022} c.f.{} their figure 3C(iii).
(b) Scatter plot of the set $\Pe(\B^\ast,\be^\ast)$ of corresponding
McKean parameters obtained by numerical solution of the inverse
problem \eqref{experiment} with calibrated $\B^\ast = 0.23487$ and
$\be^\ast= 1.85262$. 
Thin dotted lines denote the locations of the mean values of $\a_i$
and $\c_i$ with their intersection being the centre of mass $\vec\mu$ of
$\Pe$ and the violet star marker is the centre of mass
$\vec{\sigma}$ of $\Omega^\gamma_\text{ex}$.
The shaded areas are the parameter region for excitable dynamics
$\Omega^\gamma_\text{ex}$ in (b) and its pre-image in (a).
Histogram distributions with Gaussian kernel density estimations
and simple data regression lines are also plotted in both panels.}
\label{FIG050}
\end{figure*}

\looseness=-1
It is important to note that equations \eqref{experiment} are defined only on a 
subset of the full domain of excitability
$\Omega^\gamma_\text{ex}\subset \Omega_\text{ex}$ where
$\Omega_\text{ex}$ is given 
by \eqref{excitability.region}. Indeed,
for a cell to remain excitable after drug application
the  values of the parameter $\a$ must be restricted to the shorter interval
$\a\in(0,1/\be)$ rather than to the interval $(0,1)$, because after
drug application the effective
conductance is $\bar{\a}=\be\a$ and
a necessary condition for equation \eqref{APD.B} to be valid is
$\bar{\a} < 1$. Therefore,
equations \eqref{experiment} are restricted to the domain
\begin{align}
\label{excitability.region.new}
{\Omega}^\gamma_\text{ex} = \Big\{ (\a,\c) \in \mathbb{R}^2 : 
~~~ \c> -1  ~~\cap ~~
\c< \a/(1-\a) ~~\cap~~
\a>0 ~~\cap~~
\a<1/\be\Big\}.
\end{align}
The domain is visualised in Figures \ref{FIG040}(a) and \ref{FIG050}(b).

\looseness=-1
We now clarify the conditions for existence of solutions
to equations \eqref{experiment}. As noted, this system of $2N$ nonlinear algebraic
equations decouples to a set of $N$ independent pairs one per each
cell, and we consider one such pair of equations in the symbolic form
\begin{subequations}
\label{Feq}
\begin{align}
\vec{F}(\Aexp,\dAexp,\a,\c)=\vec{0},
\end{align}
where $\vec{F}:S\subset\mathbb{R}^4\to\mathbb{R}^2$ is a vector-valued
continuous function with components $F_1$
and $F_2$ given by 
\begin{align}
\vec{F}= 
\begin{pmatrix}
F_1\\
F_2
\end{pmatrix}=
\begin{pmatrix}
\Aexp/\Bexp-\A(\a,\c,\B)/\B\\
(\Aexp+\dAexp)/\Bexp-\A(\be \a,\c,\B)/\B
\end{pmatrix}.
\end{align}
\end{subequations}
We write the points in $\mathbb{R}^4$ in the form $(\vec{x},\vec{y})$ where
$\vec{x}=(\Aexp,\dAexp) \in{R}^2$ and $\vec{y}=(\a,\c)\in \mathbb{R}^2$
and recall that, by the implicit function theorem, if
$(\vec{x_0},\vec{y_0})\in S$ is a point such that
\begin{gather}
\label{conditions}
\vec{F}(\vec{x_0},\vec{y_0})=\vec{0}~~~~  \text{and}~~~~
\det\big[\partial_\vec{y} \vec{F}\big]_{(\vec{x_0},\vec{y_0})} \ne
0,
\end{gather}
\looseness=-1
then for every $\vec{x}$ in some neighbourhood of $\vec{x_0}$ there
exist a unique function with a value $\vec{y}=\vec{f}(\vec{x})$ in some
neighbourhood of $\vec{y_0}$ such that
$\vec{F}\big(\vec{x},\vec{y}=\vec{f}(\vec{x})\big)=0$. Here,
\begin{align}
\det\big[\partial_\vec{y} \vec{F}\big]= J=\det \left(\frac{\partial(F_1,F_2)}{\partial(\a,\c)}\right)
\end{align}
is the determinant of the Jacobian matrix of partial derivatives of the
components of $\vec{F}$ with respect to the components of $\vec{y}$.
Points $(\vec{x_0},\vec{y_0})$ that satisfy the first of
conditions \eqref{conditions} clearly exist because for any
$\vec{y_0}\in\Omega^\gamma_\text{ex}$ a corresponding $\vec{x_0}$ can be
computed by evaluating the right-hand-sides of
expressions \eqref{experiment}, and furthermore such pairs are unique.
Calculating the Jacobian determinant is straightforward but yields
a lengthy expression,
so to test the second of conditions \eqref{conditions} we have
evaluated it numerically and plotted the surface $J(\a,\c)$ over the
domain $\Omega^\gamma_\text{ex}$ in Figure \ref{FIG040}(a). The plot shows
that $J$ is strictly positive even though it tends to $0^+$ as
$\c\to-1$. The value $\c=-1$ is, of course, excluded from the open domain
$\Omega^\gamma_\text{ex}$ -- in particular, this is the value at which the
$\v$-nullcline and the diastolic branch of the $\u$-nullcline of the
McKean model \eqref{McKean} coincide, see discussion of
Figure \ref{fig01}(b), and it represents a non-excitable degenerate
case. 
Thus, we conclude that 
a unique solution of equation \eqref{Feq} exists in
${\Omega}^\gamma_\text{ex}$ that can be represented as
\begin{align}
\label{eq:soln}
(\a,\c)=\vec{f}\big(\Aexp,\dAexp\big),
\end{align}
and parameters can be determined as a function of experimental data $(\Aexp,\dAexp)$,

In fact, it is possible to make further progress and
find a closed form solution of equation \eqref{experiment.before} for
the parameter $\a$ as a function of $(\c, \Aexp)$ alone, yielding
\begin{align}
\label{eq:aofc}
%
\a&=1-\frac{\exp(\cnew\B)\Big(\exp(\cnew\Aexp\B/\Bexp)-1\Big)}{\cnew\Big(\exp(\cnew\B)-1\Big)\exp(\cnew\Aexp\B/\Bexp)},~~~~ \cnew =1+\c.
\end{align}
However, equation \eqref{experiment.after} is rather more difficult
to solve explicitly for $\c$. For this reason, we resort to
solving equations \eqref{experiment} numerically for both $\a$ and
$\c$. Numerical solution is straightforward: for completeness, we
mention that we have used the modified Powell hybrid
method \citep{Powell1970} as implemented in the SciPy numerical
library \citep{virtanen2020scipy}.

\subsection{Calibration of BCL and drug dose}
\label{calib}
\looseness=-1
Equations \eqref{experiment} contain, in fact, $2N+2$ unknown parameters
-- in addition to $\a_i$ and $\c_i$, the BCL $\B$ and the drug
dose parameter $\be$ to be used with the model equations are also not known \rev{a priori}. To proceed with
the analysis, unique values of $\B$ and $\be$ must be selected employing a
plausible assumption. In the absence of additional criteria, we impose the requirements 
(a) that all solutions $(\a_i,\c_i)$, $i=1,\dots, N$ belong to
the domain of excitability $\Omega^\gamma_\text{ex}$ and (b) that  these solutions are
distributed so that the distance between their centre of mass and the
centre of mass of the planar region
$\Omega^\gamma_\text{ex}$ is minimal. We will refer to this as the
``optimal excitability'' assumption as it requires that points are located
as far away as possible from the boundaries of the excitability
domain, outside of which the system is, of course, non-excitable.
Indeed, there is no prior expectation or experimental evidence that
the distribution is biased in any direction. Thus, we determine the values of $\B$ and $\be$ as
\begin{subequations}
\label{calibrateBandBe}
\begin{align}
(\B^\ast,\be^\ast) &= \arg\min\limits_{(\B,\be)}
\Big|\vec{\mu}(\B,\be)-\vec{\sigma}(\B,\be)\Big|  ~~~ \text{s.t.}~~
\Big(\a_i(\B^\ast,\be^\ast),\c_i(\B^\ast,\be^\ast)\Big) \in \Omega^\gamma_\text{ex} ~~\forall~~
i=1,\dots, N,
\end{align}
where $\vec{\mu}$ is the position vector of the centre of mass of the
set $\Pe(\B,\be)$ of discrete points $(\a_i,\c_i)$,
$i=1,\dots, N$ with coordinates in the $(\a,\c)$-plane computed as the
arithmetic means
\begin{align}
\vec{\mu}\big[\Pe(\B,\be)\big]:=\frac{1}{N}\Bigg(\sum_{i=1}^N \a_i(\B,\be), ~~~ \sum_{i=1}^N \c_i(\B,\be)\Bigg),
\end{align}
and
$\vec{\sigma}$ is the position vector of the centre of mass of the planar region
$\Omega^\gamma_\text{ex}$ which has coordinates in the $(\a,\c)$ plane
given by
\begin{align}
\vec{\sigma}\big[\Omega^\gamma_\text{ex}(\B,\be)\big]&=\left.\Bigg(\log\frac{\be}{\be-1} -\frac{1}{\be},     ~~~
\log\frac{\be-1}{\be}+\frac{1}{2(\be-1)}\Bigg)\right/\log\frac{\be}{\be-1}.
\end{align}
\end{subequations}

Criterion \eqref{calibrateBandBe} constitutes a nonlinear
minimisation problem with constraints and it is not easy to establish 
whether a unique solution to it exists. Here once again, we will restrict the
effort to a numerical demonstration instead. In Figure \ref{FIG040}(b)
contour lines of the objective minimisation surface, the distance
$\big|\vec{\mu}-\vec{\sigma}\big|$, are plotted in the $(\B,\be)$ plane
for the distribution of parameter values $(\a_i,\c_i)$ determined from
the experimental measurements $(\Aexp_i, \dAexp_i)$
of \cite{Lachaud2022}. The plot shows that the surface is globally
convex and has a single minimum, indeed. While short of a rigorous
proof, this provides an indication that a single global minimum
exists that will allow to find unique values for the model BCL $\B$
and the model drug dose parameter $\be$.

\looseness=-1
To ensure robustness, we employ
a stochastic method for
constrained global optimisation of multi-modal multi-variate objective
functions to solve the minimisation
problem \eqref{calibrateBandBe}. The method is a combination of
classical and fast
simulated-annealing approaches \citep{Tsallis1996}, coupled to a
strategy for applying a 
local search at accepted locations \citep{Xiang2000}. We use
the SciPy numerical library implementation of this dual annealing
optimisation \citep{virtanen2020scipy} which has been 
benchmarked  in \citep{Mullen2014}. Results of this approach are
presented in the next subsection. 

\subsection{Results}

Figure \ref{FIG050}(a) shows the experimentally measured data set
$\De(\Bexp,\beexp)=\big\{(\Aexp_i,\dAexp_i),~i=1,\dots, N\big\}$,
the values of which are obtained from \citep{Lachaud2022}.
Figure \ref{FIG050}(b) shows the corresponding
set of solutions  $\Pe=\big\{(\a_i,\c_i),~i=1,\dots, N\big\}$ of
the problem given by equations \eqref{experiment}
and \eqref{calibrateBandBe}. It has been obtained
numerically and is a graphical representation of the main
outcome of the approach described above.
Practically, the solution procedure involves for each pair of
$\B$ and $\be$ constructing
the domain of excitability $\Omega^\gamma_\text{ex}$ from equation \eqref{excitability.region.new}, numerically solving
the inverse problem \eqref{experiment} on this domain, computing
the distance between the respective centres of mass from
equation (\ref{calibrateBandBe}b,c), and minimising this distance with respect to $\B$
and $\be$ as per (\ref{calibrateBandBe}a). For the given experimental dataset,
this calibration procedure fixes the values of the basic cycle length and the drug
dose parameter of the model to
\begin{align}
\label{BastBEast}
\B^\ast = 0.23487,~~~~~\be^\ast= 1.85262,
\end{align}
which results in the centres of mass of $\Pe$ and
$\Omega^\gamma_\text{ex}$ located at a distance $\big|\vec{\mu}-\vec{\sigma}\big| <
1.2\times10^{-9}$ apart at
\begin{align}
\label{AastCast}
(\a^\ast,\c^\ast)  = (0.30450246, -0.24420404).
\end{align}

\begin{figure*}[t]
\raisebox{0mm}{\begin{overpic}[width=0.496\textwidth,tics=10]{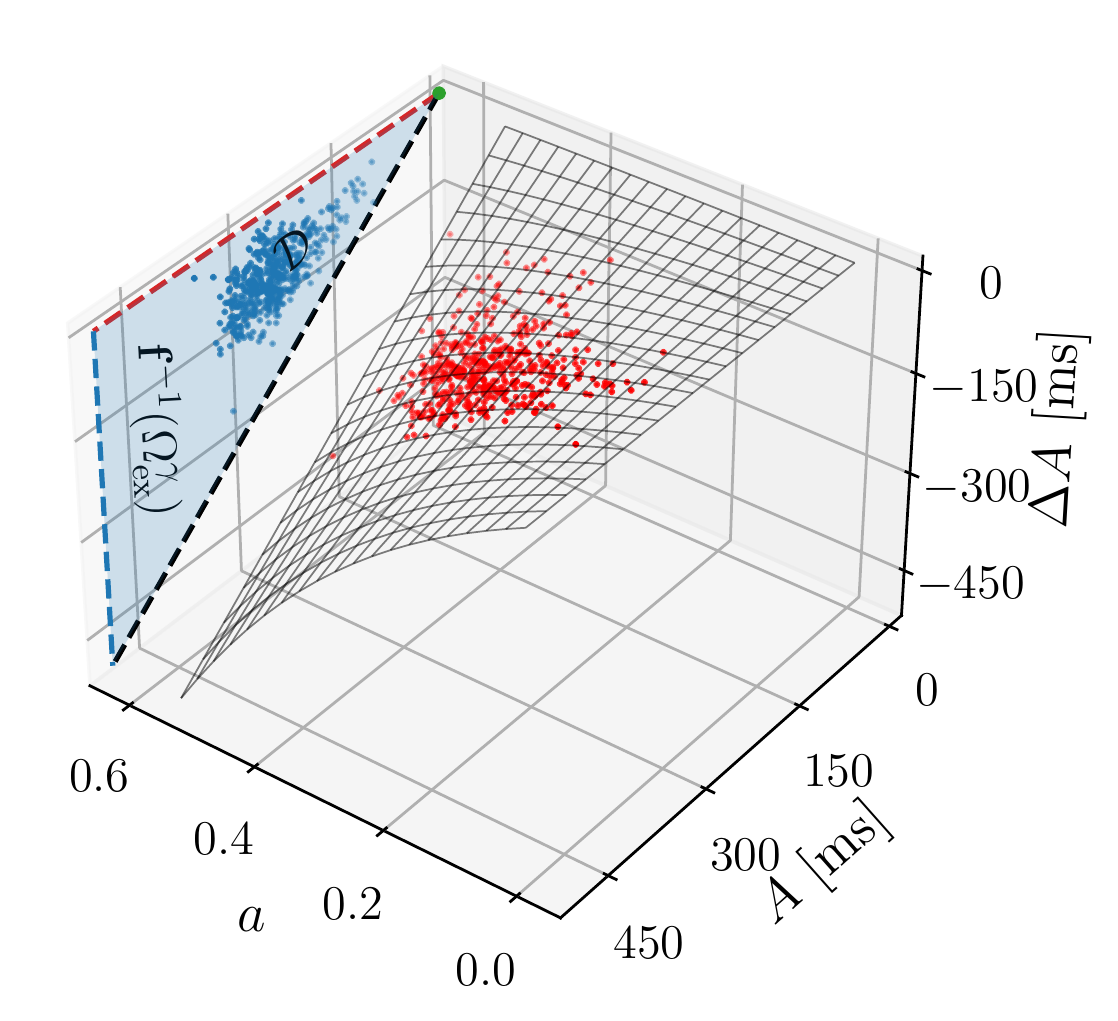}
\put (1,73) {{(a)}}
\end{overpic}}\hfill
\raisebox{2mm}{\begin{overpic}[width=0.495\textwidth,tics=10]{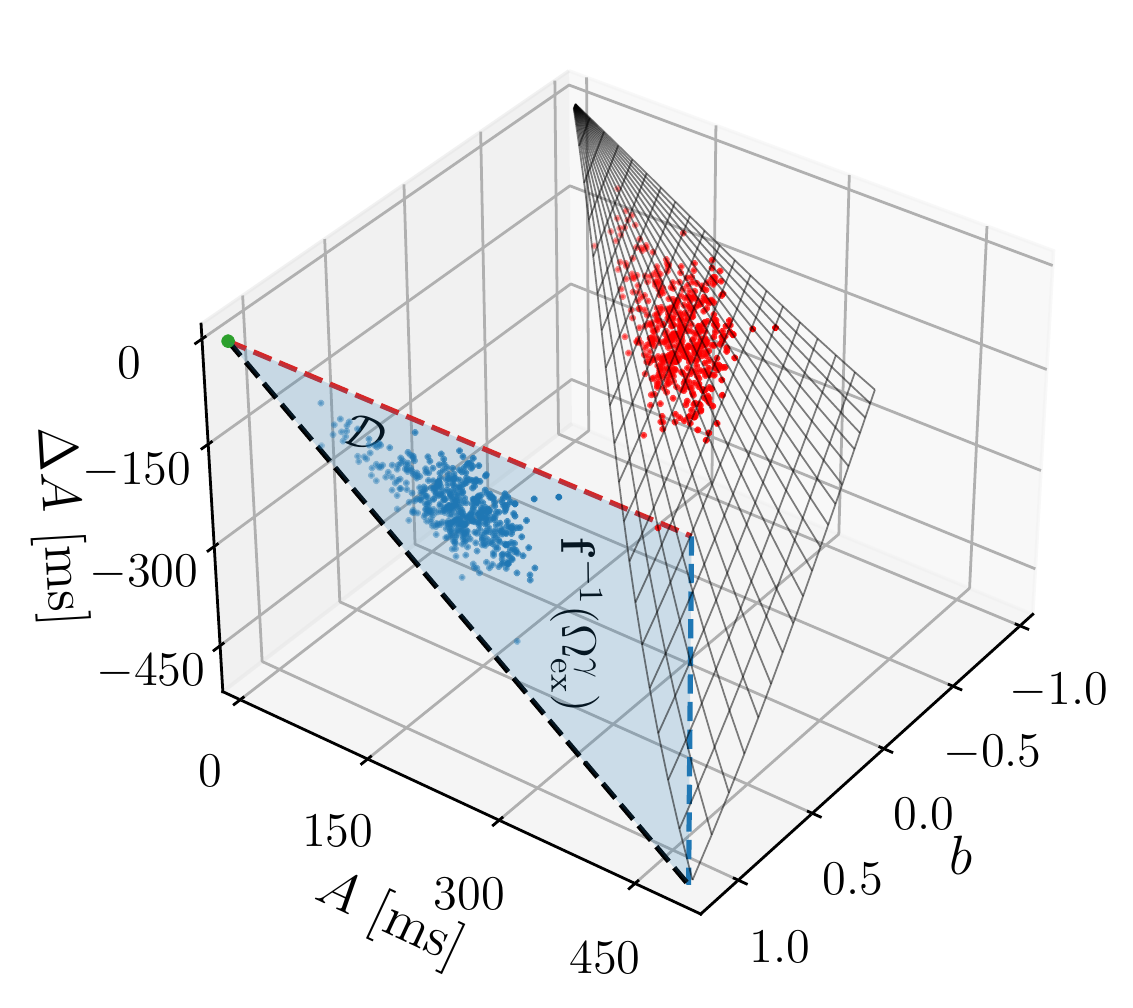}
\put (1,70) {{(b)}}
\end{overpic}}
\caption{
\looseness=-1
McKean model parameters $\a$ and $\c$ as functions of
$\Aexp$ and $\dAexp$ shown as wireframe surfaces in panel (a) and
(b), respectively. Scatter plot of the experimental data
$\De$ of \cite{Lachaud2022} is 
shown as blue dots and its projections onto the surfaces
$\a(\Aexp,\dAexp)$ and $\c(\Aexp,\dAexp)$ are shown as red dots.}
\label{FIG060}
\end{figure*}

\looseness=-1
The domain of excitability $\Omega^\gamma_\text{ex}$ in the $(\a,\c)$
plane
and its pre-image $\vec{f}^{-1}(\Omega^\gamma_\text{ex})$ in the $(\Aexp,\dAexp)$ plane, see equation \eqref{eq:soln}, are
shown in Figure \ref{FIG050}(a) and (b), respectively.
The size and shape of $\Omega^\gamma_\text{ex}$ depend on the
drug dose parameter $\be$. 
The boundaries of the image and pre-image  are
coloured correspondingly in panels (a) and (b) of Figure \ref{FIG050} to illustrate how they map into each other. It
is interesting to note that the curve $\c=-1$ maps into a single
point $(0,0)$ in the $(\Aexp,\dAexp)$ plane and from the plot of the
values of the Jacobian determinant in Figure \ref{FIG040}(a) numerical
solution of the inverse problem is expected to be most challenging in the
vicinity of this point. The relation $\vec{f}^{-1}$, which may be seen
as a change of variables transformation, maps the curves $\a=0$,
$\c=\a/(1-\a)$ and $\a=1/\be$ into the straight lines $\dAexp=0$,
$\Aexp=\Bexp$, and $\dAexp=-\Aexp$, respectively. Thus, the pre-image
$\vec{f}^{-1}(\Omega^\gamma_\text{ex})$
does not depend on $\B$ and $\be$ and 
has a triangular shape in the
$(\Aexp,\dAexp)$ plane.
This agrees with the analysis of \cite{Lachaud2022} where this region
was empirically determined, see their figure 3C(iii). It is also remarkable to observe
how well the set of experimental measurements
$\De(\Bexp,\beexp)$ is distributed and oriented within
$\vec{f}^{-1}(\Omega^\gamma_\text{ex})$.

Histograms, gaussian kernel density estimations and simple data
regression lines of the sets $\De$ and $\Pe$ are also plotted in
Figure \ref{FIG050}. Such measures are routinely used in the analysis
of experimental results, and can be compared e.g.~to figure 3C(iii)
of \citep{Lachaud2022}. Further descriptive statistical analysis of
these results can be performed but remains beyond the focus of the paper.

The map $\vec{f}$ introduced as a calibrated solution of
equations \eqref{experiment} and its inverse $\vec{f}^{-1}$ are both vector-valued functions of two arguments
$$
\mathbb{R}^2 \ni (\Aexp,\dAexp) \overset{\vec{f}}{\underset{\vec{f}^{-1}}{\myrightleftarrows{\rule{1cm}{0cm}}}} (\a,\c)\in\mathbb{R}^2.
$$
Therefore, Figure \ref{FIG050} is insufficient to identify graphically
how experimental data points in the $(\Aexp,\dAexp)$ plane map to
parameter  value points in the $(\a,\c) $ plane.
To visualise this relationship more directly, we have plotted in
Figure \ref{FIG060} the two components of $\vec{f}=(f_1,f_2)$,
namely
$\a=f_1(\Aexp,\dAexp)$ and $\c=f_2(\Aexp,\dAexp)$, separately.
This allows
\rev{to map the physiological measures into model parameters, in particular,}
to identify the value of $\a_i$ that corresponds to a given data
point $(\Aexp_i,\dAexp_i)$   from panel (a) and the value
of $\c_i$ that corresponds to the same data point $(\Aexp_i,\dAexp_i)$
from panel (b).
The surfaces $\a(\Aexp,\dAexp)$ and $\c(\Aexp,\dAexp)$ are shown as
wire-frames with grid lines that are iso-lines of the rectangular
coordinates $x$ and $y$ defined by equation \eqref{chvars} in earlier
sections.
This figure highlights the fact that an experimentally
measured value of the action potential duration $\Aexp$ is not
sufficient to characterise the cellular properties of a
myocyte. Indeed, a fixed value of $\Aexp$ corresponds to  
entire intervals as opposed to unique values for the parameters $\a$ and $\c$ 
as evident in panels (a) and (b), respectively. This point is further
discussed in the next section.

\section{Discussion and predictions} 

\subsection{Parameter interrelationships}
\begin{figure*}[t]
\raisebox{0mm}{\begin{overpic}[width=0.497\textwidth,tics=10]{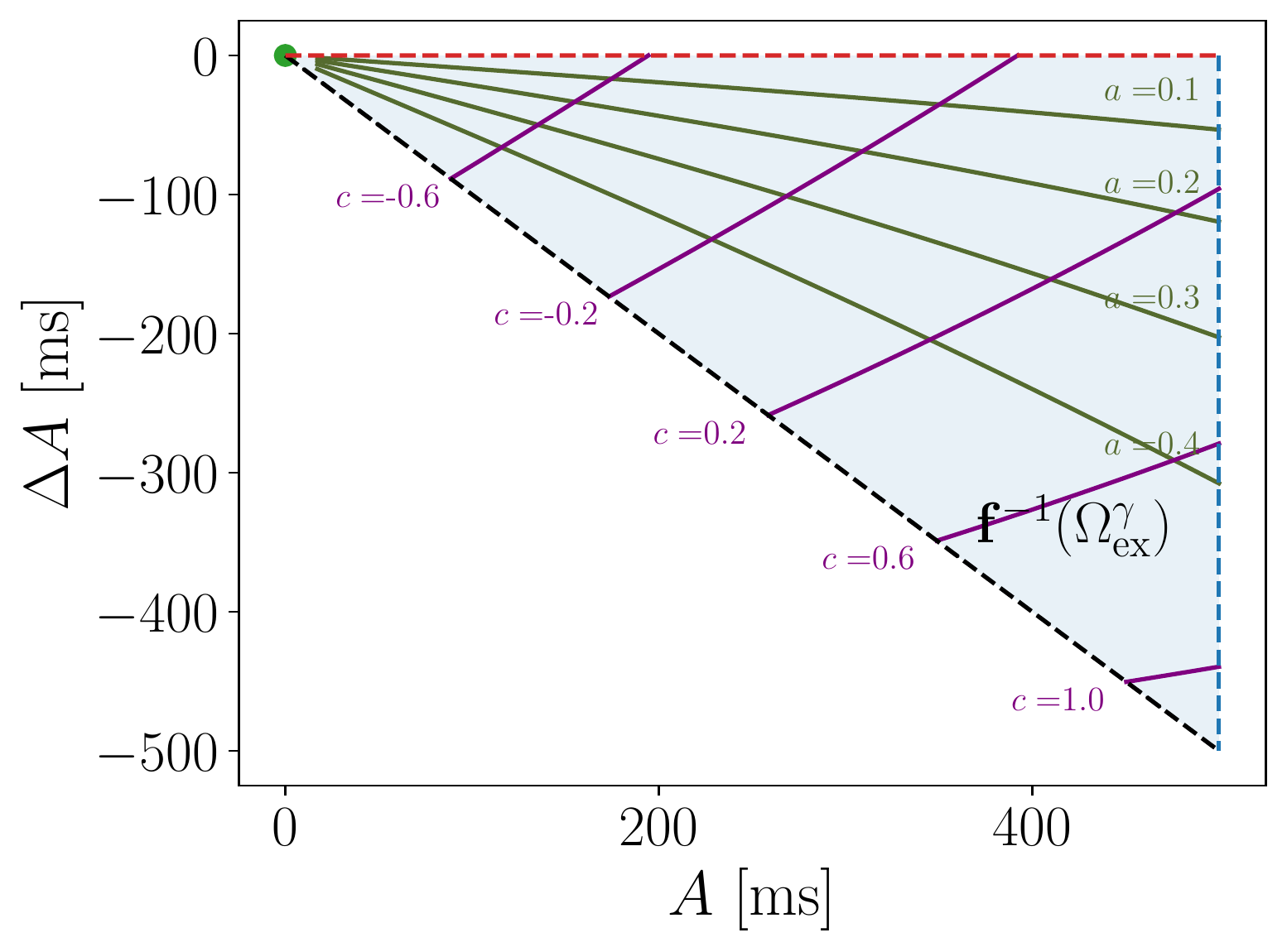}
\put (1,70) {{(a)}}
\end{overpic}}\hfill
\raisebox{0mm}{\begin{overpic}[width=0.497\textwidth,tics=10]{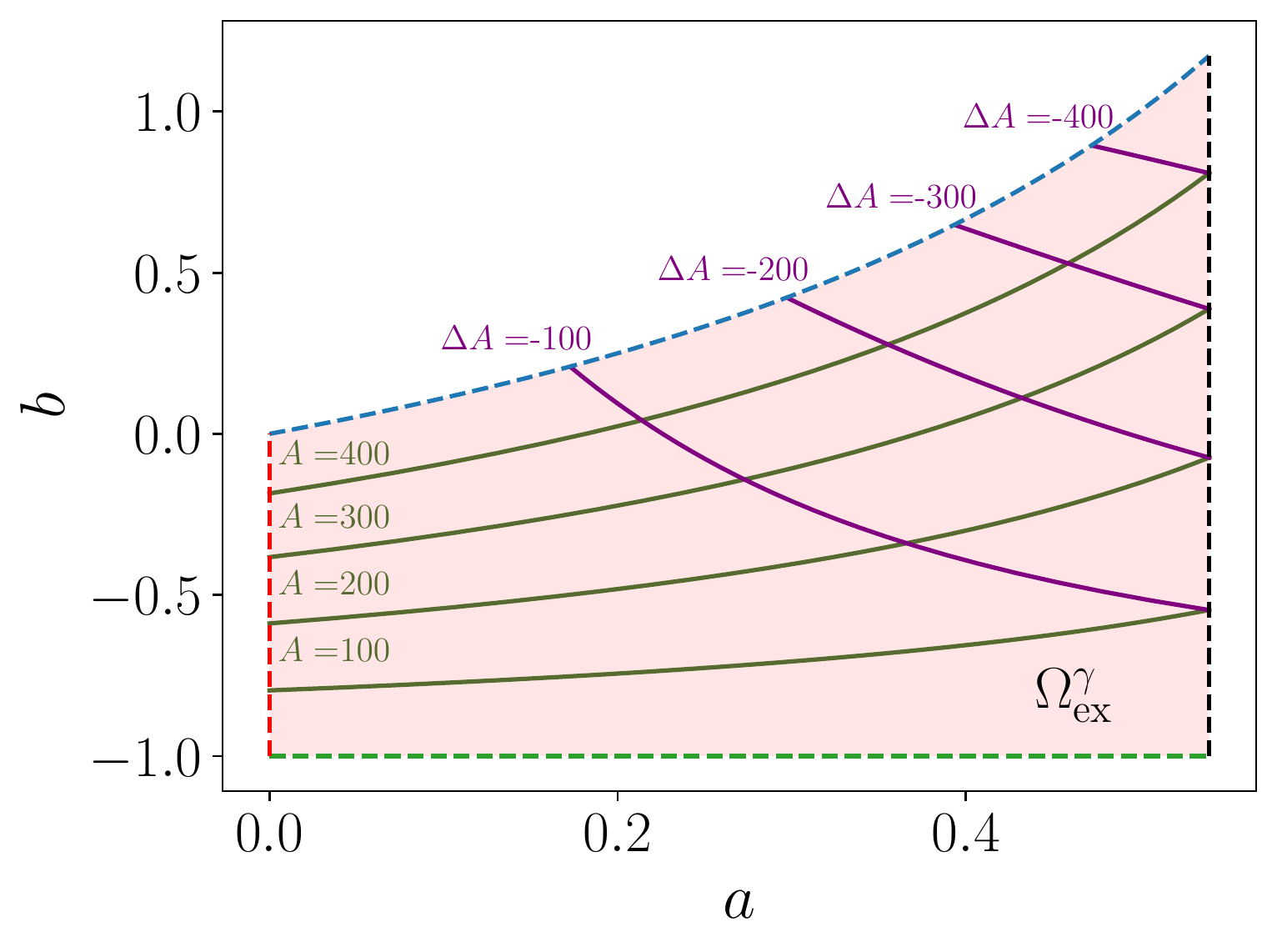}
\put (1,70) {{(b)}}
\end{overpic}}
\caption{
\looseness=-1
(a) Lines in the $(\Aexp,\dAexp)$ plane obtained at constant values of $\a$ (olive
green) and at constant values of $\c$ (purple) as labeled in the
vicinity of each curve.
(b) Lines in the $(\a,\c)$ plane obtained at constant values of $\Aexp$ (olive
green) and at constant values of $\dAexp$ (purple) as indicated in the
vicinity of each curve.
In both panels 
$\B = 0.23487$ and $\be= 1.85262$.
Other elements of the plot are similar to these described in the
caption of Figure \ref{FIG050}.
}
\label{FIG070}
\end{figure*}

The main conclusions made in the work of \cite{Lachaud2022} are (a)
``that AP morphology is retained by relationships linking specific ionic
conductances'' and (b) that ``these interrelationships are necessary for
stable repolarization despite large inter-cell variation of individual
conductances and this explains the variable sensitivity to ion channel 
block''. In an attempt to verify this assertion and to determine such
interrelationships explicitly, we have plotted a family of curves 
\begin{gather*}
\mathcal{C}_\a=\left.\big(\Aexp,\dAexp\big)\right|_{(\a=\text{const},\c)},~~~~\mathcal{C}_\c=\left.\big(\Aexp,\dAexp\big)\right|_{(\a,\c=\text{const})},
\\
\mathcal{C}_\Aexp=\left.\big(\a,\c\big)\right|_{(\Aexp=\text{const},\dAexp)},~~~~\mathcal{C}_\dAexp=\left.\big(\a,\c\big)\right|_{(\Aexp,\dAexp=\text{const})};
\end{gather*}
the first pair in Figure \ref{FIG070}(a) and the second pair in 
in Figure \ref{FIG070}(b), respectively. For example, a curve of the
type 
$\mathcal{C}_\Aexp$ represents the
interrelationship which the internal model parameters 
must satisfy so that the value of APD remains constant, and similarly
for the other types. We observe that 
the situation is more involved than suggested by \cite{Lachaud2022} in
that interrelationship represented by $\mathcal{C}_\Aexp$ changes with the particular APD value, as
well. For instance, for $\Aexp=0$ ms and small values of $\Aexp$
the dependence on $\a$ is insignificant, or there is an approximately linear
relationship between $\a$ and $\c$, while for larger values of $\Aexp$
the relationship approaches $\c=1/(1-\a)$, which is the respective
boundary of the excitability domain $\Omega^\gamma_\text{ex}$.
Similar remarks hold for the other ``grid'' lines in Figure \ref{FIG070}.
We wish to note that some of the interrelationships have already
been obtained in closed in the preceding sections. Indeed, the ``grid'' lines 
$\mathcal{C}_\a$ and $\mathcal{C}_\c$ plotted in Figure \ref{FIG070}(a) are given by the general
expression \eqref{APD.B}. Similarly, curves $\mathcal{C}_\Aexp$
and $\mathcal{C}_\dAexp$ are given by equation \eqref{eq:aofc} derived in
section \ref{existence} in relation to the conditions for solution of
problem \eqref{experiment} when evaluated for 
specified values of $\Aexp$ and $\dAexp$.

\subsection{Dependence on drug concentration and basic cycle length}
\begin{figure*}[t]
\raisebox{0mm}{\begin{overpic}[width=0.497\textwidth,tics=10]{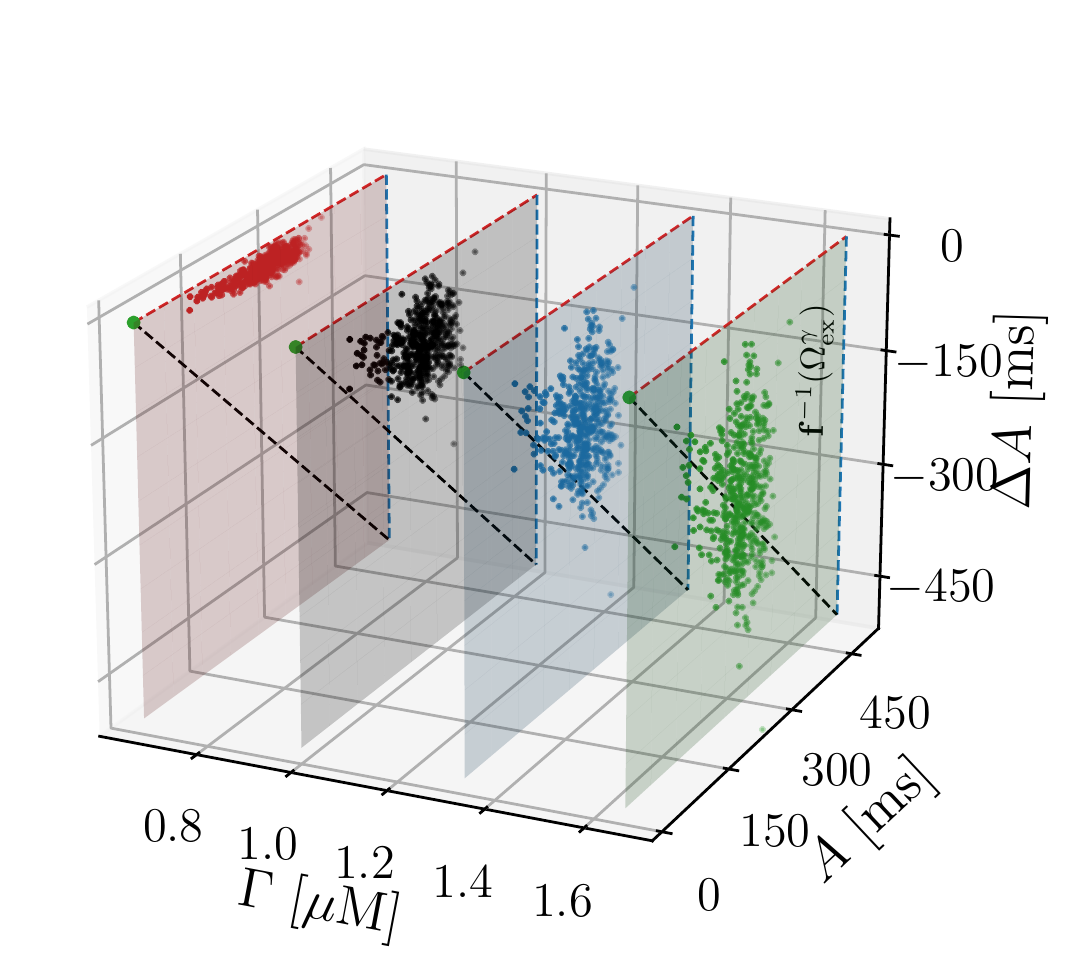}
\put (1,73) {{(a)}}
\end{overpic}}\hfill
\raisebox{0mm}{\begin{overpic}[width=0.497\textwidth,tics=10]{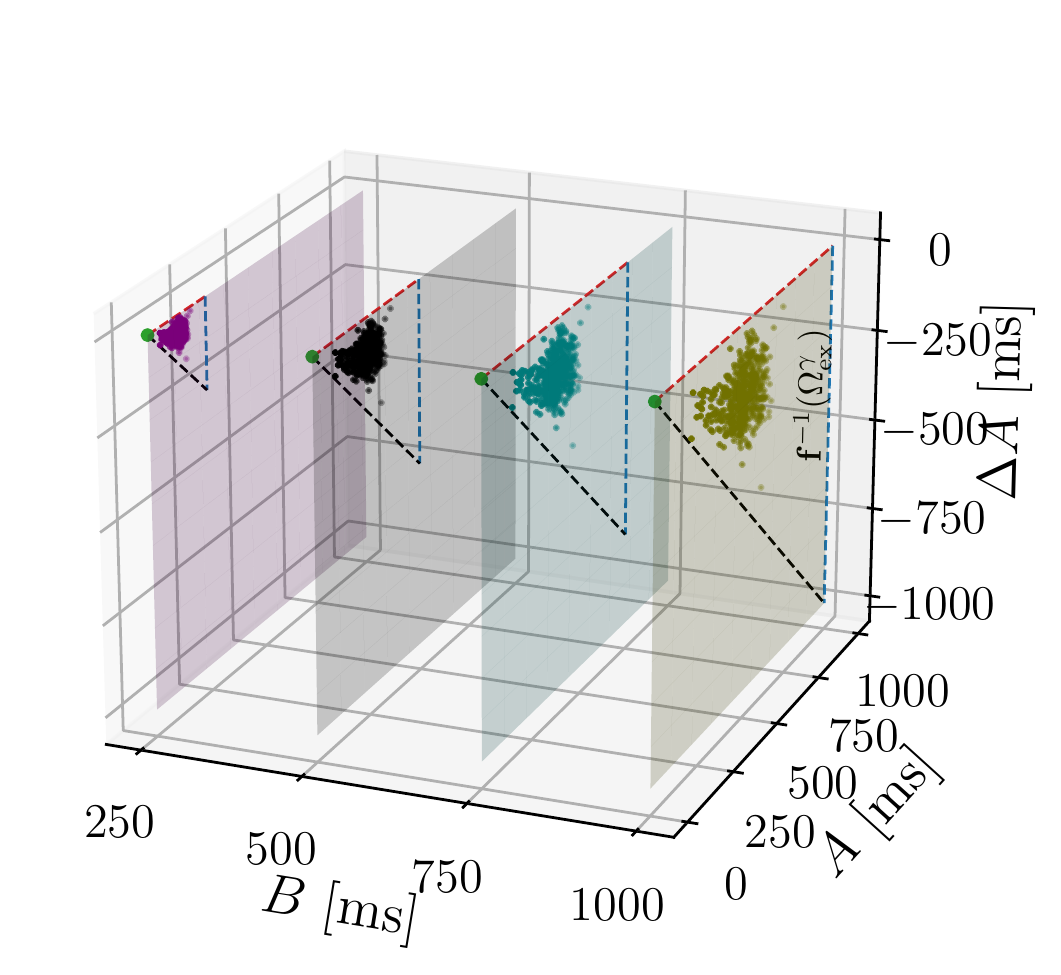}
\put (1,73) {{(b)}}
\end{overpic}}
\caption{
Prediction for the spread of the experimental data-point distribution
$\De$ of \cite{Lachaud2022} (black cloud) with variation of the drug
concentration  $\beexp$ in (a) the basic cycle length $\Bexp$ in (b).
}
\label{FIG080}
\end{figure*}

\looseness=-1
The McKean model does not offer a close correspondence to
electrophysiological myocyte structures, as conceded 
already, including in the title of the article. However, 
the strength of our approach is that, in addition to
theoretical insight, it is possible and rather economical to make
simple predictions that can be tested by (or indeed guide)
experimental measurements. Figures \ref{FIG080}(a) and (b) show predictions of
how the experimental myocyte scatter cloud $\De$ morphs when drug concentration $\beexp$ and basic 
cycle length $\Bexp$ are varied, respectively.
To produce these predictions, it is assumed that a preliminary
reference experiment at fixed values of BCL and drug concentration,
say $\Bexp^\ast$ and $\beexp^\ast$, has been performed.
The set $\Pe$ of corresponding McKean model parameter values is then
estimated by solving \eqref{experiment} and \eqref{calibrateBandBe}
with reference values $\B^\ast$ and $\be^\ast$ determined simultaneously as
discussed in section \ref{application}.
Here, for example, we continue to use the data of \cite{Lachaud2022}
which has  $\Bexp^\ast=500$ ms and $\beexp^\ast=1$ $\mu$M nifedipine
and calibration yields $\B^\ast = 0.23487$ and $\be^\ast= 1.85262$ as
before. The calculated cell-specific values $(\a_i,\c_i),~ i=1,\dots,N$, are
then used as arguments in the asymptotic expression for \APD{}
and \dAPD{}, in fact expressions \eqref{experiment} once again  
but with now varying the values of $\B$ and $\be$ away from $\B^\ast$ and $\be^\ast$. Finally, results
are converted to dimensional units using the scaling transformations
\begin{gather}
\label{eq:scaling}
\beexp=\beexp^\ast\frac{\be-1}{\be^\ast-1}, ~~~~~ \Bexp=\Bexp^\ast\frac{\B}{\B^\ast}.
\end{gather}
With decrease of the drug dose the change in action potential
duration decreases as expected. 
With increase of drug dose $\beexp$
the myocyte cloud extends down to increasingly more negative values of
$\dAexp$ 
i.e. the drug increasingly
shortens \APD{}.
The pre-image of the excitability domain does
not depend on $\beexp$, as discussed further above, and points of the myocyte cloud
eventually drift outside of it. Indeed, increasing the drug dose makes a
proportion of the cell population non-excitable. This proportion can
be estimated with respect to the reference experiment as
\begin{gather}
\label{eq:loss}
L=1-\frac{\int_{\Omega_\text{ex}^{\gamma}}dS}{\int_{\Omega_\text{ex}^{\gamma^\ast}}dS}
=\displaystyle 1- \frac{\log\big((\gamma-1)/\gamma\big)}{\log\big((\gamma^\ast-1)/\gamma^\ast\big)},
\end{gather}
where, for simplicity, we have assumed that cell properties are uniformly
distributed within their excitability domains
and the ``coefficient of loss'' $L$ is the ratio of the planar area
of $\Omega^\gamma_\text{ex}$ at concentration $\gamma$ and planar area
of $\Omega_\text{ex}^{\gamma^\ast}$ at the reference concentration $\gamma^\ast$.
Equation \eqref{eq:loss} can be re-cast in terms  of drug
concentration $\beexp$ with the help of the change of variables \eqref{eq:scaling}.
With variation of the basic cycle length $\Bexp$
the myocyte cloud seems to undergo a
shape-preserving scaling transformation - it enlarges with increase of
$\Bexp$ or shrinks with decrease of $\Bexp$ while keeping shape as
seen in Figure \ref{FIG080}(b). The pre-image
$\vec{f}^{-1}(\Omega^\gamma_\text{ex})$ does change size in a similar way
with $\Bexp$. The myocyte cloud scales non-linearly as determined
by equation \eqref{APD.B} and illustrated in the restitution curve of
Figure \ref{FIG030}(b) and would thus ``saturate'' for values of
$\Bexp$ larger than shown in  Figure \ref{FIG080}(b).

\begin{figure*}[t]
\raisebox{0mm}{\begin{overpic}[width=0.497\textwidth,tics=10]{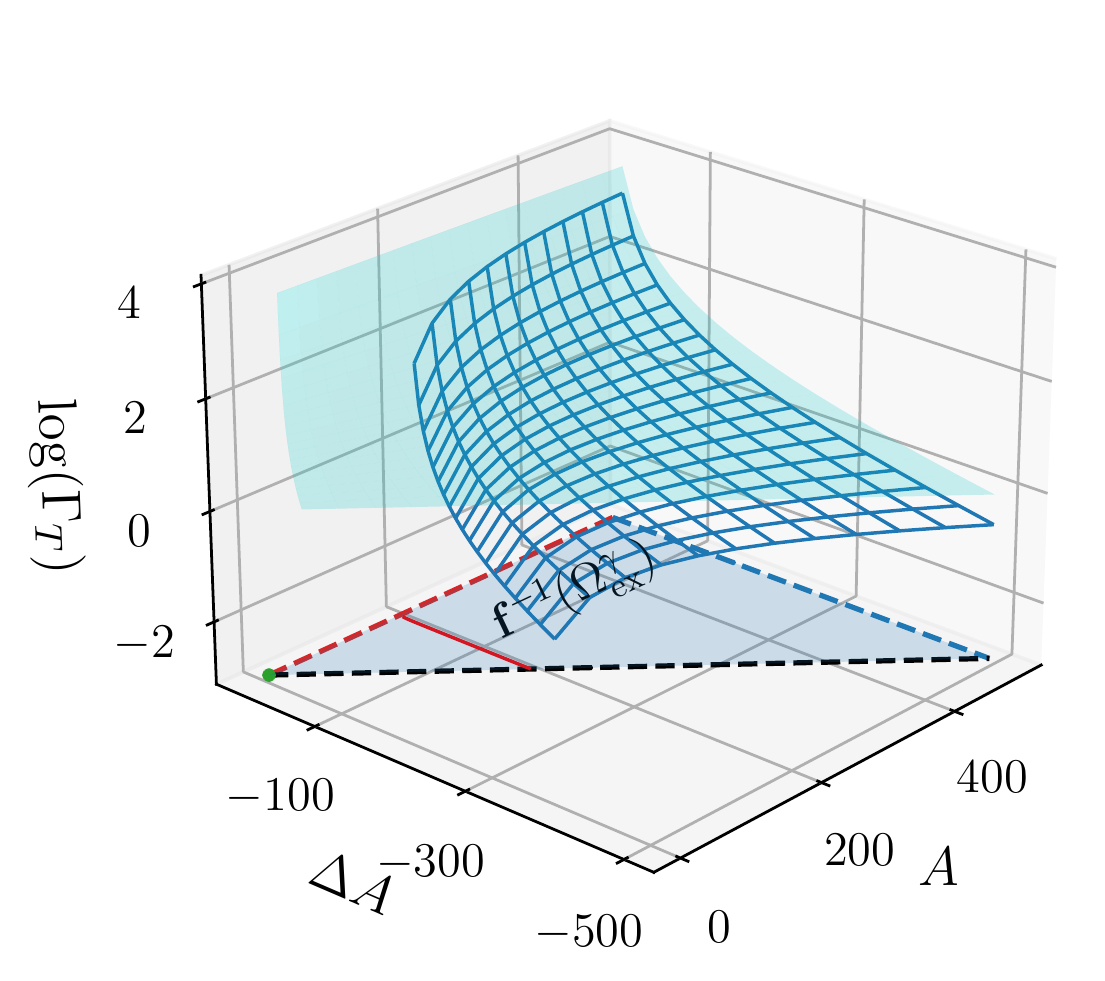}
\put (1,73) {{(a)}}
\end{overpic}}\hfill
\raisebox{2mm}{\begin{overpic}[width=0.497\textwidth,tics=10]{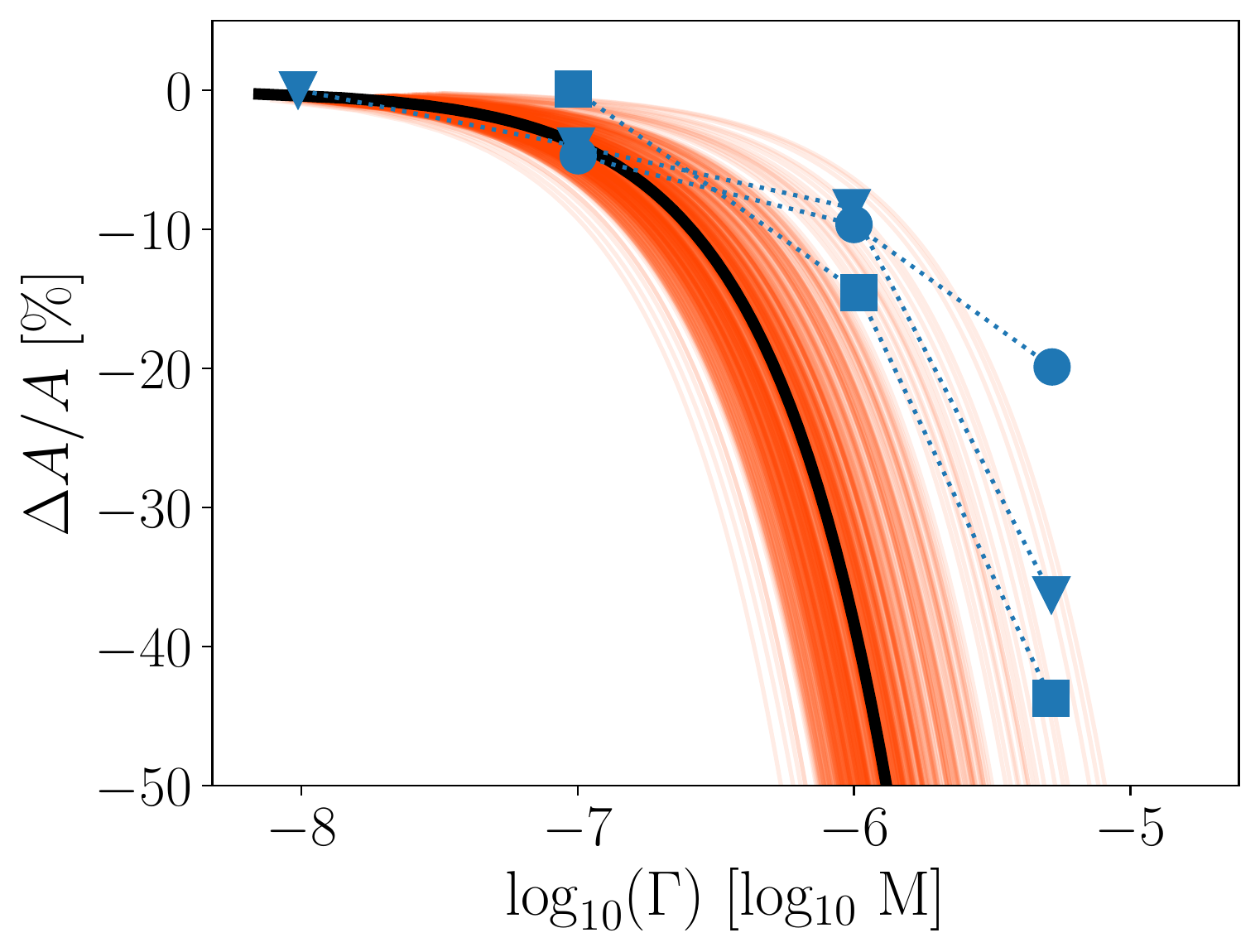}
\put (1,73) {{(b)}}
\end{overpic}}
\caption{\looseness=-1
(a) Values $\beexp_T$ of the drug concentration (nifedipine in $\mu$M)
necessary to elicit action potentials with a prescribed duration
$\Aexp_T=180$ ms in a heterogeneous population of myocytes (blue
wireframe surface). 
The solid red line is a projection of the line $\beexp_T=0$ $\mu$M and the
transparent aquamarine surface is $\beexp_T=\beexp^\ast\,(1-\a)/(\be^\ast-1)/\a$, both constraining the
acceptable values $\beexp_T$ can take. (b) Dose-response 
curve of nifedipine computed using \eqref{experiment.after}. Thin semi-transparent lines are dose-response
curves for individual cells in the myocyte population
of \cite{Lachaud2022}, with the solid black line corresponding to
the ``mean cell'' with parameter values given
in \eqref{AastCast}. Experimental data (blue dotted lines) from rabbit
Purkinje fibre action potentials (triangles down) and ventricular action
potentials (circles) and field potentials (squares) in thin-slice
tissue preparations is also shown (data from figure 6B
of \citep{Himmel2012}). In both panels, McKean parameters $(\a,\c)$
are calibrated to experimental data of \cite{Lachaud2022} using
$\Bexp^\ast=500$ ms, $\beexp^\ast = 1$ $\mu$M nif.~and $\B^\ast = 0.23487$ and
$\be^\ast= 1.85262$.
}
\label{FIG090}
\end{figure*}

\subsection{Dose-response curves}
\looseness=-1
Perhaps most significant from a control and medical intervention
viewpoint is the problem to determine a ``target'' value of the drug
concentration, $\beexp_T$  say, to be administered to a population of myocytes such
that all myocytes respond with an  identical
target ``healthy'' action potential under periodic stimulation. We
will denote the duration of this target AP by $\Aexp_T$. The target concentration $\beexp_T$ sought will
be different for each cell in the population. Assuming as
before that a preliminary reference experiment has been performed at
fixed $\Bexp^\ast$ and $\beexp^\ast$, that $\B^\ast$ and $\be^\ast$
have been calibrated and the McKean parameters have been determined
then the target drug concentration $\beexp_{T}$ is given by
\begin{align}
\label{eq:target}
\beexp_T= \frac{\beexp^\ast}{\be^\ast-1} \left(\frac{1}{\a}-\frac{\exp(\cnew\B^\ast)\Big(\exp(\cnew\Aexp_T\B^\ast/\Bexp^\ast)-1\Big)}{\a\cnew\Big(\exp(\cnew\B^\ast)-1\Big)\exp(\cnew\Aexp_T\B^\ast/\Bexp^\ast)}-1\right),~~~~ \cnew =1+\c.
\end{align}
The latter is, of course, precisely expression \eqref{eq:aofc} where the target
value of the McKean ``conductance'' parameter is related to the one
determined in the reference experiment by
$\a_T=\be_T \a$ and scaling \eqref{eq:scaling} to dimensional values
has been applied. Further, values of $\be_T$ determined from this
expression must be additionally subjected to the (a) constraints $\be_T
>1$ or equivalently $\beexp_T>0$ $\mu$M nif and (b)
$\be_T<1/\a$. Constraint (a) reflects the fact that the minimal drug
concentration that can be administered is 0 $\mu$M, and constraint (b) is
the, now familiar, requirement that for a cell to be excitable $\a\be_T
<1$ must hold.
Figure \ref{FIG090}(a) illustrates these results and
shows the target drug concentration values for nifedipine in $\mu$M 
with values $\B^\ast$ and $\be^\ast$ calibrated for the data
of \cite{Lachaud2022}. The surface $\beexp_T$, logarithmically
transformed for clarity of visualisation, is plotted both as a function
of the experimental biomarkers of the controlled experiment
$(\Aexp,\dAexp)$ and a similar plot can be constructed in terms of corresponding McKean model parameters
$(\a,\c)$.  While constraint (b) appears to be always satisfied, constraint (a)
restricts the range of myocytes for which the drug intervention can
work. This is easy to understand -- as nifedipine is an APD shortening
drug, for cells with APD already shorter than the target APD, there is
no amount of nifedipine that can be administered to prolong APD to the
target value. A second drug with a different mechanism may, of course,
be deployed.

\looseness=-1
It is, of course, not possible to administer different drug dose to
each individual cell. Figure \ref{FIG090}(b) provides more 
conventional dose-response curves computed using
equation \eqref{experiment.after} for each cell in the population as 
well as a for the ``mean cell'' in the population, the point with
parameter values given in equation \eqref{AastCast}.
These
predicted dose-response curves are also compared in the Figure to
experimental data from \citep{Himmel2012} where the dose response to
nifedipine was measured by the changes elicited in action potentials
and field potentials in thin slices of rabbit ventricular tissue
and rabbit Purkinje fibres. 
The results show a remarkable agreement,
given \rev{there are very significant differences in experimental
configuration (tissue slices as opposed uncoupled cells), types of
measurement (field potentials in some cases) and the
essential variation in cellular properties (Purkinje as opposed to
ventricular myocytes, also unrelated animals) that is now known to exist
between populations.} The latter heterogeneity has been the underlying
motivation for 
the present work.

\section{Conclusion}

In this work, a simple conceptual model of cellular
excitability is employed to analyse experimental measurements of 
ion channel block in a large and heterogeneous population of uncoupled
cardiomyocytes. The experimental data, due to \cite{Lachaud2022},
consists principally of values of the action potential duration
shortening measured in nearly 500 rabbit ventricular myocytes to which
1 $\mu$M of the drug nifedipine was applied. The cells were sourced form various regions
of the left ventricles of several different animals and exhibited
a significant intrinsic variation in their action potential duration and
drug response, already quantified by \cite{Lachaud2022}. The
main aim of our analysis is to infer the cellular properties of each
myocyte in terms of cell-specific parameter values of an appropriate
mathematical model and the \cite{McKean1970} model was selected for
this  purpose due to its simplicity. The \cite{McKean1970} model is a
fast-slow system of piece-wise linear ordinary differential equations
of FitzHugh-Nagumo type. It has two
variables that can be interpreted as voltage and an effective gating
variable and two intrinsic parameters that can be seen as
an effective ion current conductance and an effective kinetic
parameter. Here, the domain in parameter space where the model
exhibits excitable dynamics (as opposed to oscillatory or bistable) is
determined, and an asymptotic approximation of the duration of 1:1
action potentials generated by strictly periodic stimulation is
obtained using a  standard fast-slow asymptotic
analysis \citep{Tikhonov-1952,Fenichel-1979}. The
approximation takes the form of an explicit analytical expression for
the APD as a function of the \cite{McKean1970} model parameters and
the basic cycle length of stimulation, i.e. $\A(\a,\c,\B)$. Such a
relation is known as a restitution curve/relation in the electrophysiological
literature. The drug action of nifedipine is modelled by introducing
a multiplicative factor $\be$ to the effective conductance parameter $\a$,
yielding a problem for the solution of a set of non-linear algebraic
equations from which the McKean model parameters $\a_i$ and $\c_i$ for
each cell $i=1,\dots, N$, can be determined given experimental
measurements of the action potential durations $\Aexp_i$ and
$\Aexp_i+\dAexp_i$ recorded under periodic stimulation with basic
cycle length $\Bexp$ before and after drug application with concentration $\beexp$. Remarkably,
this results in an adaptive domain problem, where the 
parameter domain $\Omega^\gamma_\text{ex}(\B,\be)$, the 
basic cycle length $\B$ and the drug dose parameter $\be$
must be determined as a part of the solution.
This is done by introducing a further modelling assumption that the
euclidean distance between its centroid of the domain and
the algebraic mean of the McKean parameter values of the population is
minimal. It is demonstrated by direct numerical
evaluation that (a) the adaptive domain minimisation and (b) the set of
2N nonlinear algebraic equations both admit unique solutions, that are
then found using standard numerical routines.
\rev{In particular, the existence of $\Omega^\gamma_\text{ex}(\B,\be)$
different from $\Omega_\text{ex}(\B,\be)$, suggests that
when implementing heterogeneity in realistic models, parameter values
should be selected from a restricted region of the parameter space only.}
The results are then also used (a) to understand interrelationships
proposed by \cite{Lachaud2022} as necessary to ensure generation of
stable AP morphology and repolarisation, (b) \rev{to predict} the scatter of
APD values of the population with variation of basic cycle length and
drug concentration,  (c) to calculate nifedipine drug-response curves
for each cell in the population and determine a value for the drug concentration
required so that \rev{each uncoupled cell, and the population as a whole,
responds with a single APD value}, and (d) \red{predict} the proportion of cells that become
inexcitable at large drug doses. \rev{Prediction} (c) is found to compare
well with independent experimental measurements \citep{Himmel2012},
while the other predictions may also be tested or indeed guide experimental measurements. 

The methodology presented here can be extended, refined and applied in a number of
directions. (a) Being piece-wise linear, the \cite{McKean1970} model
allows exact solutions in closed form and thus the asymptotic
expression \eqref{APD.B} used here may be replaced by more a more
accurate exact expression. This will then allow to incorporate
experimental measurements of secondary AP biomarkers, such as action
potential duration at 50\% and 30\% from peak, impossible to
distinguish using the present asymptotic expression. (b) In place of
the \cite{McKean1970}
model, the caricature Noble model proposed
by \cite{Biktashev-2008} as an archetypal model of cardiac
excitability may be used. It has the advantage of having been derived by a controlled and
systematic procedure form an actual ionic current model and captures
the fundamental mathematical structure of cardiac electrical
excitability. \rev{In particular, it includes a super-fast subsystem,
lacking in the \cite{McKean1970} model, which will allow analysis of biomarkers describing the front (Phase
1) of the action potential, such as the time from 10\% to 90\% of
upstroke (TRise) which has been also measured experimentally and depends on
cellular processes essentially different from these that control
action potential duration.} The caricature Noble model has been
recently fitted to reproduce the action potential morphology and
restitution properties of several cardiomyocyte
phenotypes \citep{Aziz2022} and has known exact and asymptotic
solutions \citep{Biktashev-2008,Simitev2011}, albeit more involved.
(c) In our work only the  drug action on the effective conductance parameter
$\a$ is considered, and the attention is further restricted to the case
of action potential shortening as needed for comparison to the
nifedipine dataset of \cite{Lachaud2022}. The cases of action
potential prolongation and of drug action applied to the
effective kinetic parameter $\c$ need to be investigated, as well. Action
potential prolongation is induced for instance by the drug dofetilide
with measurements also reported in  \citep{Lachaud2022}. In this case
the parameter domain of excitability must be adapted in a different
way meriting a separate investigation.
(d) It will be of interest to extend the current methodology to
include coupling between myocytes and thus investigate AP waveform
synchronisation. This is likely to play a significant role in generation a stable action
potential response on tissue-wide level. These are all directions open for
future research.

\paragraph{Acknowledgements}
This work was supported by the UK Engineering and Physical Sciences
Research Council [grant numbers EP/S030875/1 and EP/T017899/1]. 

\setlength{\bibsep}{1.0mm}
\bibliographystyle{apalike3}
\bibliography{refs}

\end{document}